\documentclass[journal=jctcce,manuscript=article]{achemso}

\usepackage[version=3]{mhchem} 
\usepackage{array} 

\author{Sriram Anand}
\affiliation{Department of Materials Science and Engineering, McMaster University, 1280 Main Street West, Hamilton, Ontario L8S 4L8, Canada}
\alsoaffiliation{Department of Metallurgical and Materials Engineering, National Institute of Technology Tiruchirappalli, Tamil Nadu 620015, India}
\author{Caio Miranda Miliante}
\affiliation{Department of Materials Science and Engineering, McMaster University, 1280 Main Street West, Hamilton, Ontario L8S 4L8, Canada}
\author{Storm Gourley}
\affiliation{Department of Chemical Engineering, McMaster University, 1280 Main Street West, Hamilton, Ontario L8S 4L8, Canada}
\author{Brian D. Adams}
\affiliation{Salient Energy Inc., Dartmouth, Nova Scotia B3B 1C4, Canada}
\author{Drew Higgins}
\affiliation{Department of Chemical Engineering, McMaster University, 1280 Main Street West, Hamilton, Ontario L8S 4L8, Canada}
\author{Oleg Rubel}
\email{rubelo@mcmaster.ca}
\affiliation{Department of Materials Science and Engineering, McMaster University, 1280 Main Street West, Hamilton, Ontario L8S 4L8, Canada}

\title
  {Computational screening of cathode materials for Zn-ion rechargeable batteries}

\begin{document}

\begin{tocentry}

\includegraphics{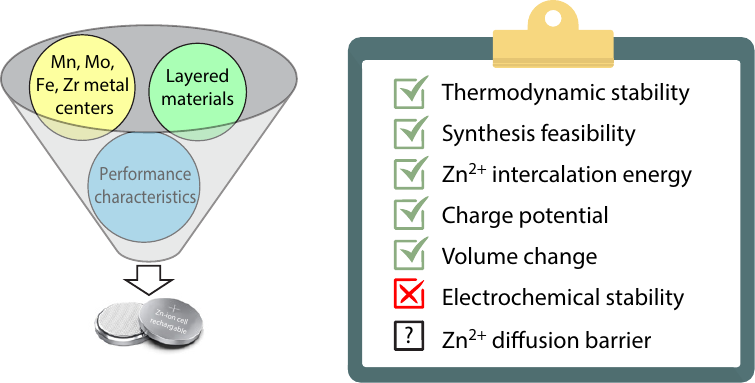}

\end{tocentry}

\begin{abstract}
We propose a comprehensive set of indicators (including methods to obtain and analyse them) for computational screening of candidate cathode materials for rechargeable Zn-ion aqueous batteries relying on \ce{Zn^{2+}} intercalation processes. The indicators capture feasibility of \ce{Zn^{2+}} intercalation and transport within the material, the thermodynamic stability of charged and discharged material structures, electrochemical stability of the cathode material and electrolyte, volume expansion, and energy storage capacity. The approach was applied to well-known cathode materials (\ce{\alpha-MnO2} and \ce{\alpha-V2O5}) as well as several potential alternatives (\ce{MoS2}, \ce{ZrP2O7}, \ce{MoO3}, and \ce{FeO2}) to demonstrate the screening workflow and the decision making process. We show that selection of cathode materials for Zn-ion aqueous rechargeable batteries is a multifaceted problem, and first principle calculations can help to narrow down the search. Despite us being unable to identify a particularly successful cathode material, tools and techniques developed in this work can be applied more broadly to screen a wider array of potential material compositions and structures, with the goal of identifying next generation cathode materials for aqueous rechargeable batteries with the intercalation energy storage mechanism not limited to \ce{Zn^{2+}} ions.
\end{abstract}

\section{Introduction}\label{sec:Introduction}

Owing to their light-weight, high energy density and long-term cyclability, lithium ion (Li-ion) batteries have become front-runners for energy storage in handheld devices and electric vehicles. However, the scarcity of raw materials reduces their economic feasibility, and the flammability and toxicity of the electrolytes used adds further concerns of operational safety \cite{Cabana_AM_22_2010,Blanc_J_4_2020}. Recently, the focus of battery research has broadened towards multivalent ion batteries, such as \ce{Zn^{2+}}, \ce{Mg^{2+}} and \ce{Al^{3+}} \cite{Fan_AM_30_2018,Pan_AEM_11_2021}.

Rechargeable aqueous zinc-ion batteries (RAZIBs) offer several benefits in terms of the balance of performance, cost, environmental impact and material abundance, and consequently stand apart from current battery technologies \cite{Zhang_CSR_49_2020}. Two major bottlenecks to the progress and commercial realization of RAZIBs are the electrochemical stability of the cathode during the intercalation-deintercalation cycles and risk of electrolyte oxidation (i.e., the oxygen evolution reaction) \cite{Selvakumaran_JMCA_7_2019}.

Oxides of manganese have been previously studied \cite{Zhao_I_2_2019,Pan_NE_1_2016,Chen_AMI_6_2019,Jiang_EA_229_2017,Selvakumaran_JMCA_7_2019} for application as RAZIB cathodes due to their abundance, ease of processing and eco-friendly nature. The main concern of RAZIBs is the electrochemical stability of the cathode material (dissolution of \ce{Mn^{2+}} into the electrolyte and co-deposition of non-stoichiometric manganese oxides) that is responsible for capacity fade over time \cite{Pan_NE_1_2016,Selvakumaran_JMCA_7_2019,Wu_EES_13_2020,Blanc_J_4_2020,Tran_SR_11_2021,Rubel_JPCC_126_2022}. Vanadium based cathodes gained recognition for their excellent capacity retention, especially the layered \ce{\alpha-V2O5} cathode, however, these materials are more expensive and as cathodes produce low discharge voltages relative to \ce{MnO2}, making them less favourable in applications that require high voltages and energy density \cite{Blanc_J_4_2020}. Prussian blue analogues are another prominent class of cathodes with tested discharge potentials of ca.~1.7~V. However, for the purpose of this study, they are not considered due to their low specific capacities \cite{Selvakumaran_JMCA_7_2019}. Discovery of cathode materials that mitigate such drawbacks is hence of great importance for large-scale realization of RAZIBs.

Previously, cathode materials for Li-ion batteries \cite{Chevrier_PRB_82_2010,Aydinol_PRB_56_1997} and  multivalent ion batteries \cite{Zhang_AO_4_2019} have been modeled to determine their discharge potentials by estimating the energy associated with the working ion insertion using the density functional theory (DFT) \cite{Hohenberg_PR_136_1964,Kohn_PR_140_1965}. A Battery Explorer application in the Materials Project \cite{Jain_AM_1_2013} allows screening of cathode materials candidates using the charge potential and several complementary indicators, such as the theoretical specific capacity, the volume change as well as stability of charged and discharged materials relative to the energy convex hull. Feasibility of kinetic processes within the cathode material can be indirectly evaluated by determining the diffusion barrier for the charged ion migration also with DFT \cite{Ong_EES_4_2011}. However, the electrochemical stability of cathode materials remains overlooked in existing screening protocols.

Exploration and screening of suitable RAZIB cathode materials can also be expedited by providing theoretical estimates of thermodynamic characteristics associated with \ce{Zn^{2+}} intercalation. To date computational efforts in the field of RAZIB cathode materials are scarce. \citet{Le_AAMI_13_2021} reported calculations of the \ce{Zn^{2+}} deintercalation potential in \ce{\alpha-MnO2} showing that hydration along with the Hubbard $U$ correction in DFT are important factors for reproducing the experimental intercalation potential. Our group showed that it is possible to reproduce the experimental deintercalation potential in electrolytic \ce{MnO2} at the PBE+$U$ level of theory using the spinel (\ce{$\lambda$-Zn_{0-1}Mn2O4}) structure as a model without hydration \cite{Rubel_JPCC_126_2022}. (Here PBE stands for \citet*{Perdew_PRL_77_1996} exchange-correlation functional.) \citet{Luo_CCL_unknown_2022} recently reported high-throughput screening of spinel materials with a general formula \ce{$XY$2O4} (where $X$ and $Y$ are metals with one of them being either Mn or Zn) as cathodes in RAZIB. Their screening parameters were the \ce{Zn^{2+}} intercalation potential, the volume change upon charge/discharge, the \ce{Zn^{2+}} diffusion coefficient, the thermodynamic stability of the \ce{$XY$2O4} structure \cite{Luo_CCL_unknown_2022}, and zero band gap as an indicator of good electrical conductivity. Shortcomings of Ref.~\citenum{Luo_CCL_unknown_2022} are the omission of the Hubbard $U$ correction leading to a mismatch between calculated and measured charge potentials, unrealistic spinel-derived \ce{Zn_{$x$}$XY$2O4} structures, and omission of the electrochemical stability analysis. This leaves a room for development of a comprehensive procedure for screening of RAZIB cathode materials using thermodynamic parameters computed with DFT or extracted from literature, which can adequately determine their real-world applicability.

In this paper, we used \ce{Zn/MnO2} and \ce{Zn/V2O5} as well-studied rechargeable zinc-ion battery materials to verify and benchmark the predictive power of first-principle calculations. We propose a screening protocol that includes analysis of the structural features to allow kinetic insertion and extraction of \ce{Zn^{2+}} ions, \ce{Zn^{2+}} deintercalation potential, thermodynamic stability of the cathode in charged and discharge states, feasibility of synthesis of the pristine cathode material, energy storage capacity, volume change due to Zn intercalation, and electrochemical stability. We demonstrated the approach by applying the protocol to several potential cathode materials for RAZIBs with different redox active metal centers: \ce{FeO2}, \ce{MoS2}, \ce{ZrP2O7}, and \ce{MoO3}. With this procedure we aim to significantly diminish the time for evaluating the characteristics of a potential cathode material, as it takes not more than one week to analyze a relatively simple structure (up to 50 atoms), while the experimental route spans any time between a few weeks to a few months.

\section{Computational details}\label{sec:Method}

DFT calculations were performed using the Vienna \textit{ab initio} simulation package (VASP) \cite{Kresse_PRB_47_1993,Kresse_CMS_6_1996,Kresse_PRB_54_1996}. The generalized gradient approximation (GGA) with the Perdew-Burke-Ernzerhof (PBE) \cite{Perdew_PRL_77_1996} parametrization was used for approximating the exchange-correlation energy  along with the DFT-D3 method with Becke-Johnson damping \cite{Grimme_JCP_132_2010,Grimme_JCC_32_2011} to capture van der Waals interactions. Projector augmented wave pseudopotentials \cite{Kresse_PRB_59_1999} were employed in all calculations. The Brillouin zone was sampled using a Monkhorst-Pack grid of $k$ points \cite{Monkhorst_PRB_13_1976} generated automatically with a linear density of 20 divisions per 1~{\AA}$^{-1}$ of the reciprocal space. The plane wave cut-off energy specified by the pseudopotentials was increased by 25\% to achieve an accurate stress tensor. The energy convergence threshold was set at $10^{-5}$~eV. We performed full structure relaxations for all compounds. The relaxation threshold for forces acting on atoms was set at 0.05~eV~{\AA}$^{-1}$ and for the stresses at 1~kbar. The initial magnetic ordering was assumed ferromagnetic with the exception of Mn compounds, where we explored alternative antiferromagnetic spin configurations. The effect of magnetic ordering (ferromagnetic vs antiferromagnetic) has a minor effect on total energies. For instance, the DFT PBE+D3+$U$ total energy per formula unit (f.u.) of \ce{$\alpha$-MnO2} changes from $-21.509$~eV/f.u. (antiferromagnetic) to $-21.510$~eV/f.u. (ferromagnetic).

To account for correlations effects the PBE exchange-correlation was augmented with a Hubbard $U$ correction (a simplified approach introduced by \citet{Dudarev_PRB_57_1998}). The performance of DFT+$U$ for modelling transition-metal oxides and electrochemical reactions has been shown to be the more accurate method owing to better representation of localised valence \textit{d} and \textit{f} electrons \cite{CapdevilaCortada_AC_6_2016}. Here we adopted $U_{\text{eff}}$ values benchmarked in Ref.~\citenum{Jain_PRB_84_2011} for the transition metal (TM) elements listed in Table~\ref{tab:U} throughout all compounds where they are present.

\begin{table}
\caption{Hubbard PBE+$U$ parameters for $d$ electrons of TM species taken from Ref.~\citenum{Jain_PRB_84_2011}. The single parameter $U_\text{eff}=U-J$ is used according to \citet{Dudarev_PRB_57_1998} DFT+$U$ approach.}
\label{tab:U}
\begin{tabular}{ l @{\qquad} c }
    \hline
    Element & $U_\text{eff}$ (eV) \\
    \hline
    Mn & 3.9\\
    V & 3.1\\
    Fe & 4.0\\
    Mo & 3.5\\
    \hline
\end{tabular}
\end{table}

A nudged elastic band (NEB) method \cite{Jonsson_class-quant-dynam_1998, Mills_SS_324_1995} was used to calculate the \ce{Zn} diffusion energy barrier between its favored intercalation position in a unit cell for the benchmark materials, \ce{$\alpha$-V2O5} and \ce{\alpha-MnO2}. VASP's implementation of the method \cite{Jonsson_class-quant-dynam_1998, Mills_SS_324_1995} was used with the spring constant set to $-5$ and 16 intermediate `images' between the initial and final points. To improve convergence of the NEB path for \ce{\alpha-V2O5}, the search for a minimum energy paths was initially executed without the Hubbard $U$ correction. Then, the calculated images were used as a starting point for the final calculation with the Hubbard $U$ correction.

The central thermodynamic quantity for analysis of stability of materials is the formation energy. It is calculated (in eV per f.u.) using the expression
\begin{equation}\label{Eq:dH_f}
    \Delta H_\text{f} = H(\text{compound}) - \sum_i^{\text{elements}} N_{i}\, H(i, \text{bulk}),
\end{equation}
where the summation index $i$ runs over all constituting elements of the compound, $H$ stands for DFT total energies (per f.u. for the compound or per atom for individual elements), and $N$ is the number of atoms $i$ in the compound (per f.u.). The elemental solids (or molecules) are considered in their most stable form. The pressure times volume ($PV$) enthalpy term is ignored. It only reaches a sizable magnitude for \ce{O2} gas still contributing only 0.023~eV per molecule, which is an order of magnitude smaller than the chemical accuracy of PBE (see Table~\ref{tab:delH}).

\begin{table}
\caption{Formation energies (eV/f.u.) of cathode materials computed using Eq.~\eqref{Eq:dH_f} compared with experimental enthalpies and other calculations reported in literature.}
\label{tab:delH}
\begin{tabular}{ l c c c l }
    \hline
    Material & PBE+D3 & PBE+D3+$U$ & Experimental & Calculated (literature) \\
    \hline
    \ce{$\beta$-MnO2} & $-5.07$ & $-4.66$ & $-5.39$ \cite{LandoltBornstein_thermo_2001} & $-4.78$ (PBE) \cite{Franchini_PRB_75_2007}\\
    &&&& $-5.07$ (PBE) \cite{Eckhoff_PRB_101_2020}\\
    &&&& $-4.63$ (PBE+$U=3$~eV) \cite{Franchini_PRB_75_2007}\\
    \ce{\alpha-MnO2} & $-5.02$ & $-4.69$ & --- & $-5.42$ (ID: mp-19395) \cite{Jain_AM_1_2013} \\
    \ce{$\lambda$-MnO2} & $-4.85$ & $-4.56$ & --- & $-4.91$ \cite{Eckhoff_PRB_101_2020}\\
    \ce{\alpha-V2O5} & $-15.62$ & $-15.73$ & $-16.07$ \cite{Chase_JPCRD_Monograph9_1998} & $-16.11$ (PBE) \cite{Das_CMS_163_2019}\\
    &&&& $-16.03$ (PBE+D3+$U=3.5$~eV) \cite{Das_CMS_163_2019}\\
    \ce{$\alpha$-FeO2} & --- & $-3.85$ & --- & $-3.73$ (ID: mp-796324) \cite{Jain_AM_1_2013} \\
    \ce{$\lambda$-FeO2} & --- & $-3.37$ & --- & $-3.31$ (ID: mp-540003) \cite{Jain_AM_1_2013} \\
    \ce{MoO3} & --- & $-7.95$ & $-7.72$ \cite{Chase_JPCRD_Monograph9_1998} & $-7.54$ (ID: mp-20589) \cite{Jain_AM_1_2013}\\
    \ce{MoS2} & --- & $-2.52$ & $-2.86$ \cite{Chase_JPCRD_Monograph9_1998} & $-3.61$ (ID: mp-2815) \cite{Jain_AM_1_2013}\\
    \ce{ZrP2O7} & $-27.0$  & --- & $-27.4$ \cite{brown2005chemical} & $-31.07$ (ID: mp-5024) \cite{Jain_AM_1_2013} \\
    \hline
\end{tabular}
\end{table}

The insertion of Zn should be energetically favorable for a cathode material to be a viable candidate. We assess this by calculating the corresponding energy (in eV per Zn atom)
\begin{equation}\label{eq:intercalation energy}
    \Delta H_{\text{Zn}} =
    \frac{H(\ce{Zn}_{x}M_{y}\ce{O}_{z}) - H(M_{y}\ce{O}_{z}) - x H(\ce{Zn})}{x}
\end{equation}
based on PBE+D3+$U$ total energies $H$. Negative value of $\Delta H_{\text{Zn}}$ indicates favorable Zn insertion.

The \ce{Zn^{2+}} deintercalation potential of a \ce{Zn$_{x}M_{y}$O$_{z}$} compound is obtained from the Nernst equation
\begin{equation}\label{eq:deintercalation potential}
    E \approx -\,\frac{\Delta H_{\text{Zn}}}{n_{\text{e}} e},
\end{equation}
where $n_{\text{e}}$ is the number of electrons transferred per ion (2 for \ce{Zn^{2+}}) and $e$ is the elementary charge. This expression is equivalent to that originally proposed by \citet{Aydinol_PRB_56_1997}. Here we neglect finite temperature effects assuming that they largely cancel out thus the approximate sign in Eq.~\eqref{eq:deintercalation potential}.

The theoretical specific capacities of a cathode material \ce{Zn$_{0 \ldots x}M_{y}$O$_{z}$} was computed using the relation \cite{Wu_PNSMI_29_2019}
\begin{equation}\label{Eq:Specific_Capacity}
    C = \frac{x n_{\text{e}} F}{3.6 M_w},
\end{equation}
where $F$ is the Faraday constant and $M_w$ is the molar mass of the host compound \ce{$M_{y}$O$_{z}$}. The factor 3.6 accounts for units conversion to yield mA~h~g$^{-1}$.

\section{\label{sec:Results and Discussion}Results and Discussion}

\subsection{Validation of methods using \ce{Zn/V2O5} and \ce{Zn/MnO2}}

\begin{figure}
    \begin{center}
        \includegraphics{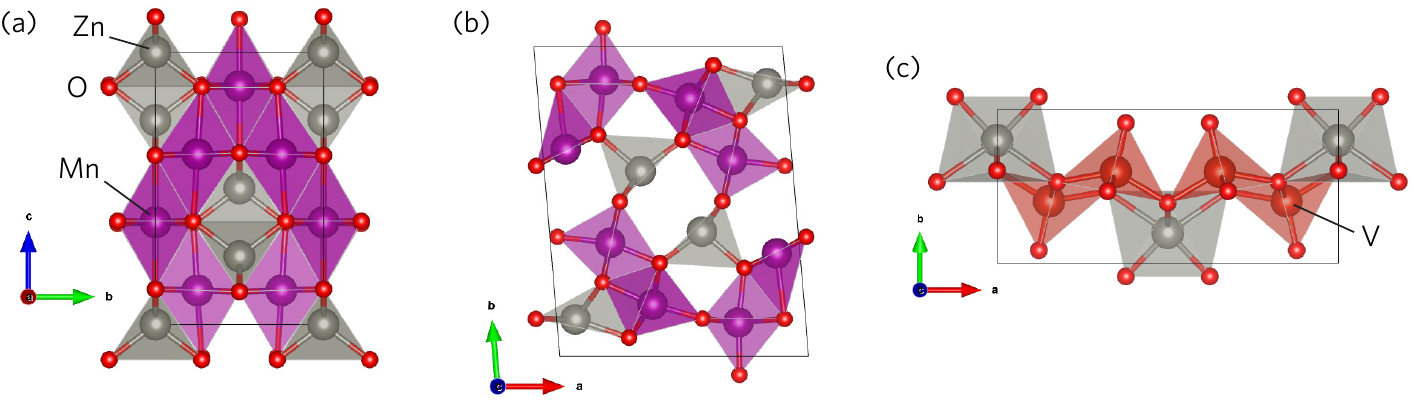}
    \end{center}
    \caption{Benchmark structures with intercalated Zn: (a) spinel \ce{ZnMn2O4}, (b) hollandite-like \ce{ZnMn2O4}, and (c) layered \ce{ZnV2O5}.}
    \label{fig:struct-benchmark}
\end{figure}

\ce{Zn/MnO2} and \ce{Zn/V2O5} systems are taken as benchmarks for the purpose of this study owing to prior experimental works on their use as RAZIB cathodes~\cite{Xu_ACIE_51_2011,Alfaruqi_CM_27_2015,Kundu_NE_1_2016,Song_AFM_28_2018,Yang_JMCA_8_2020}. Crystal structures corresponding to a discharged state (with intercalated Zn) are presented in Fig.~\ref{fig:struct-benchmark}. In addition to the standard spinel \ce{ZnMn2O4} phase, the structure of hollandite-like \ce{ZnMn2O4} was derived from \ce{\alpha-MnO2} by placing Zn inside tunnels. Multiple alternative Zn arrangements were tested in attempt to achieve an optimal coordination of Zn atoms. The lowest energy structure is presented in Fig.~\ref{fig:struct-benchmark}(b), which is similar to the one suggested computationally in Ref.~\citenum{Le_AAMI_13_2021}. It should be noted, however, that there is no experimentally resolved  hollandite-like \ce{ZnMn2O4} structure reported in the literature. Similarly, there is no native \ce{ZnV2O5} structure. The derived structure of \ce{ZnV2O5} (Fig.~\ref{fig:struct-benchmark}c) was inspired by the experimental atomic-resolution studies of Zn insertion in \ce{\alpha-V2O5} \cite{Byeon_NC_12_2021}.

Calculated structures of materials used in the benchmark study (\ce{$\alpha,\lambda$-MnO2}, spinel \ce{ZnMn2O4}, and \ce{\alpha-V2O5}) are compared to experiment in Table~\ref{tab:latparam}. Structures computed using PBE+D3 exchange-correlation functional have an average error of 0.2\% in lattice parameters (1.2\% maximum absolute error), while PBE+D3+$U$ achieves a greater average error of 0.8\% in lattice parameters (2.2\% maximum absolute error). This level of accuracy is within limits expected for DFT with semilocal exchange-correlation functionals \cite{Haas_PRB_79_2009}.

\begin{table}
\caption{Calculated lattice parameters of materials compared with experimental data from a Springer Materials database \cite{Villars_SpringerMaterials_2016} (a unique database identifier is given in brackets).}
\label{tab:latparam}
\begin{tabular}{ l c l l }
    \hline
    Material & Space group & Method & Lattice parameters ({\AA})\\
    \hline
    \ce{\alpha-MnO2} & I4/m (87) & PBE+D3 & $a=9.75,c=2.86$ \\
    && PBE+D3+$U$ & $a=9.82,c=2.91$ \\
    && Exp. (sd\_0542154) & $a=9.815,c=2.847$\\
    \hline
    \ce{$\lambda$-MnO2} & Fd$\bar{3}$m (227) & PBE+D3 & $a=8.09$ \\
    && PBE+D3+$U$ & $a=8.18$ \\
    && Exp. (sd\_1142474) & $a=8.060$\\
    \hline
    \ce{ZnMn2O4} & I4$_1$/amd (141) & PBE+D3 & $a=5.73,c=9.29$ \\
    && PBE+D3+$U$ & $a=5.78,c=9.34$ \\
    && Exp. (sd\_0377570) & $a=5.772,c=9.236$\\
    \hline
    \ce{\alpha-V2O5} & Pmmn (59) & PBE+D3 & $a=11.65,b=3.57,c=4.37$ \\
    && PBE+D3+$U$ & $a=11.73,b=3.62,c=4.28$ \\
    && Exp. (sd\_0313359) & $a=11.510,b=3.563,c=4.369$\\
    \hline
    \ce{MoO3} & Pnma (62) & PBE+D3+$U$ & $a=14.71,b=3.72,c=3.88$ \\
    && Exp. (sd\_0530932) & $a=14.02,b=3.703,c=3.966$\\
    \hline
    \ce{MoS2} & P6$_3$/mmc (194) & PBE+D3+$U$ & $a=3.17,c=12.24$ \\
    && Exp. (sd\_0309036) & $a=3.16,c=12.29$\\
    \hline
    \ce{ZrP2O7} & Pa$\bar{3}$ (205) & PBE+D3+$U$ & $a=8.50$ \\
    && Exp. (sd\_1222320) & $a=8.272$\\
    \hline
\end{tabular}
\end{table}

Chemical accuracy of DFT calculations was tested by computing the formation energies $\Delta H_\text{f}$ of benchmarked compounds and comparing them with experiments as well as calculated values reported in literature (Table~\ref{tab:delH}). Both choices for the exchange-correlation functional (PBE+D3 and PBE+D3+$U$) systematically underestimate $\Delta H_\text{f}$ for oxides which is consistent with numerous prior studies \cite{Franchini_PRB_75_2007,Jain_PRB_84_2011,Eckhoff_PRB_101_2020}. Even though the addition of a Hubbard correction $U$ does not lead to improving $\Delta H_\text{f}$, however, it does improve the description of correlation effects in localized orbitals of TM oxides and allows to get oxidation energies (and, therefore, redox energies) correctly \cite{Wang_PRB_73_2006} which becomes important later when calculating a \ce{Zn^{2+}} deintercalation potential and charge voltage.

With formation energies at hand, we can verify stability of intercalated structures. However, it is not sufficient to claim stability based on the negative formation energy alone. It is important to perform a energy convex hull analysis and verify that the material in question (\ce{Zn$_{x}M_{y}$O$_{z}$}) is \textit{on} the convex hull. For this purpose, we plot a section of the ternary convex hull \ce{Zn-$M_{y}$O$_{z}$}. Assuming that \ce{$M_{y}$O$_{z}$} is stable, we can define  a relative stability parameter $\delta H$ (in eV per atom)
\begin{equation}\label{Eq:Rel_stability}
\begin{split}
\delta H & = \frac{H(\ce{Zn}_{x}M_{y}\ce{O}_{z}) - H(M_{y}\ce{O}_{z}) - x H(\ce{Zn})}{x+y+z} \\
 & = \frac{\Delta H_f(\ce{Zn$_{x}M_{y}$O$_{z}$}) - \Delta H_f(\ce{$M_{y}$O$_{z}$})}{x+y+z} .
\end{split}
\end{equation}
Here we replaced the total energies $H$ with formation energies $\Delta H$ since such values are also available experimentally. Negative values of $\delta H$ suggest that Zn insertion is favorable. The \ce{Zn$_{x}M_{y}$O$_{z}$} compound is considered stable if it has the lowest $\delta H$ among other polymorphs of the same compound or combination of other compounds with equivalent composition, which makes this analysis analogous of the energy convex hull.

\begin{figure}
    \includegraphics{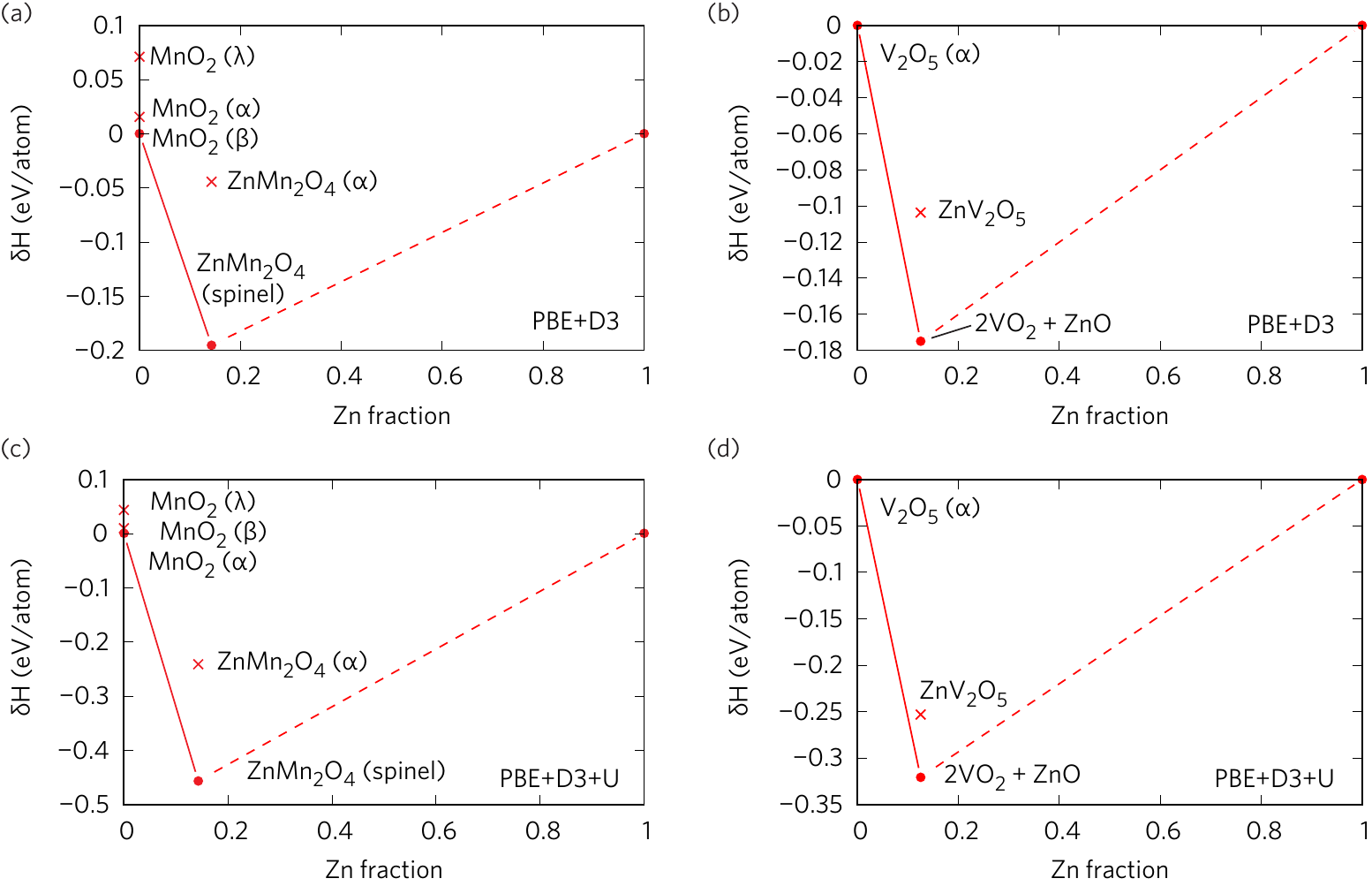}
    \caption{\ce{MnO2-Zn} and \ce{V2O5-Zn} energy convex hull diagrams: (a,b) computed at the PBE+D3 level of theory, (c,d) computed at the PBE+D3+$U$ level of theory.}
    \label{fig:deltaH-MnO2-V2O5}
\end{figure}

The relative stability parameter $\delta H$ is presented in Fig.~\ref{fig:deltaH-MnO2-V2O5}.  The spinel \ce{ZnMn2O4} is the most stable structure being on the convex hull, while hollandite-like \ce{ZnMn2O4} and \ce{ZnV2O5} are metastable since they are above the convex hull with yet negative $\delta H$. Results with and without the Hubbard $U$ correction are qualitatively the same. Quantitatively, PBE+D3+$U$ results in a more discern convex hull and significantly more negative $\delta H$ making Zn intercalation even more favorable.

The \ce{Zn^{2+}} deintercalation potential vs \ce{Zn^{0}/Zn^{2+}} calculated according Eq.~\eqref{eq:deintercalation potential} for the benchmark structures in Fig.~\ref{fig:struct-benchmark} are listed in Table~\ref{tab:Zn-potential-benchmark}. Calculations without the Hubbard $U$ correction systematically underestimate the \ce{Zn^{2+}} deintercalation potential. Inclusion of the Hubbard $U$ correction is essential to reproduce experimental deintercalation potentials as also evident from prior studies of cathode materials for Li-ion batteries  \cite{Chevrier_PRB_82_2010,Urban_nCM_2_2016} since PBE exchange-correlation functional systematically underestimates the intercalation and redox potentials \cite{Jain_PRB_84_2011,Chevrier_PRB_82_2010}. The correction compensates inaccuracy of the semi-local PBE exchange-correlation potential and lowers energy of $d$ states at the TM ion. As a result, calculations reproduce more accurately an energy gained by the transfer of electrons from Zn to the TM ion. For this reason PBE+D3+$U$ will be used to evaluate alternative cathode materials for the remainder of this paper.

\begin{table}
\caption{\ce{Zn^{2+}} deintercalation potentials (V) vs \ce{Zn^{0}/Zn^{2+}} for benchmark structures in Fig.~\ref{fig:struct-benchmark} computed with and without the Hubbard correction $U$. The average experimental Zn extration and insertion potential are listed for comparison.}
\label{tab:Zn-potential-benchmark}
\begin{tabular}{ l c c c c }
    \hline
    Cathode & TM redox couple & PBE+D3 & PBE+D3+$U$ & Experimental \\
    \hline
     $\lambda$-\ce{MnO2} & \ce{Mn^{4+/3+}} & 0.9 & 1.7 & $1.5 \pm 0.5$ \cite{Wu_JMCA_5_2017,Tang_EMA_2022_2022} \\
     $\alpha$-\ce{MnO2} & \ce{Mn^{4+/3+}} & 0.3 & 1.1 & 1.5 \cite{Xu_ACIE_51_2011,Alfaruqi_JPS_288_2015,Lee_CC_51_2015} \\
     \ce{\alpha-V2O5} & \ce{V^{5+/4+}} & 0.4 &  1.0 & 0.8 \cite{Kundu_NE_1_2016,Pang_ASS_538_2021}, 1.0 \cite{Hu_AAMI_9_2017} \\
     \hline
\end{tabular}
\end{table}

Despite applying the $U$ correction, we are unable to fully recover the experimental potential for \ce{\alpha-MnO2} (Table~\ref{tab:Zn-potential-benchmark}). We attempted to include hydration of Zn ions to achieve their proper coordination within the \ce{\alpha-MnO2} structural voids, but it did not improve the agreement with experiment (see Sec.~\ref{sec:results:role-hydration}). Other mechanisms contributing to the energy storage (\ce{MnO2} dissolution or \ce{H^+} intercalation) can mask \ce{Zn^{2+}} intercalation as they occur within the same potential window \cite{Lee_SR_4_2014,Wu_EES_13_2020,Moon_AS_8_2021,Rubel_JPCC_126_2022,Chen_AM_34_2022}. Thus, the spinel phase gives a better representation of \ce{ZnMn2O4} thermodynamics at DFT level than its hollandite-like counterpart. There is other evidence suggesting structural instabilities of highly loaded \ce{\alpha-MnO2}: Large distortions of $\alpha$-like \ce{K_{$x$}MnO2}  ($x>0.25$) was reported by \citet{Jiao_JMCA_8_2020} who attributed it to a transformation from a tunnel to a layered structure (we can see that in Fig.~\ref{fig:struct-benchmark}(b) as well). \citet{Le_AAMI_13_2021} in DFT studies observed structural transformation and transition from $\alpha$ to a layered $\delta$ structure starting at \ce{Zn_{$0.375$}MnO2}.

\begin{figure}
    \includegraphics{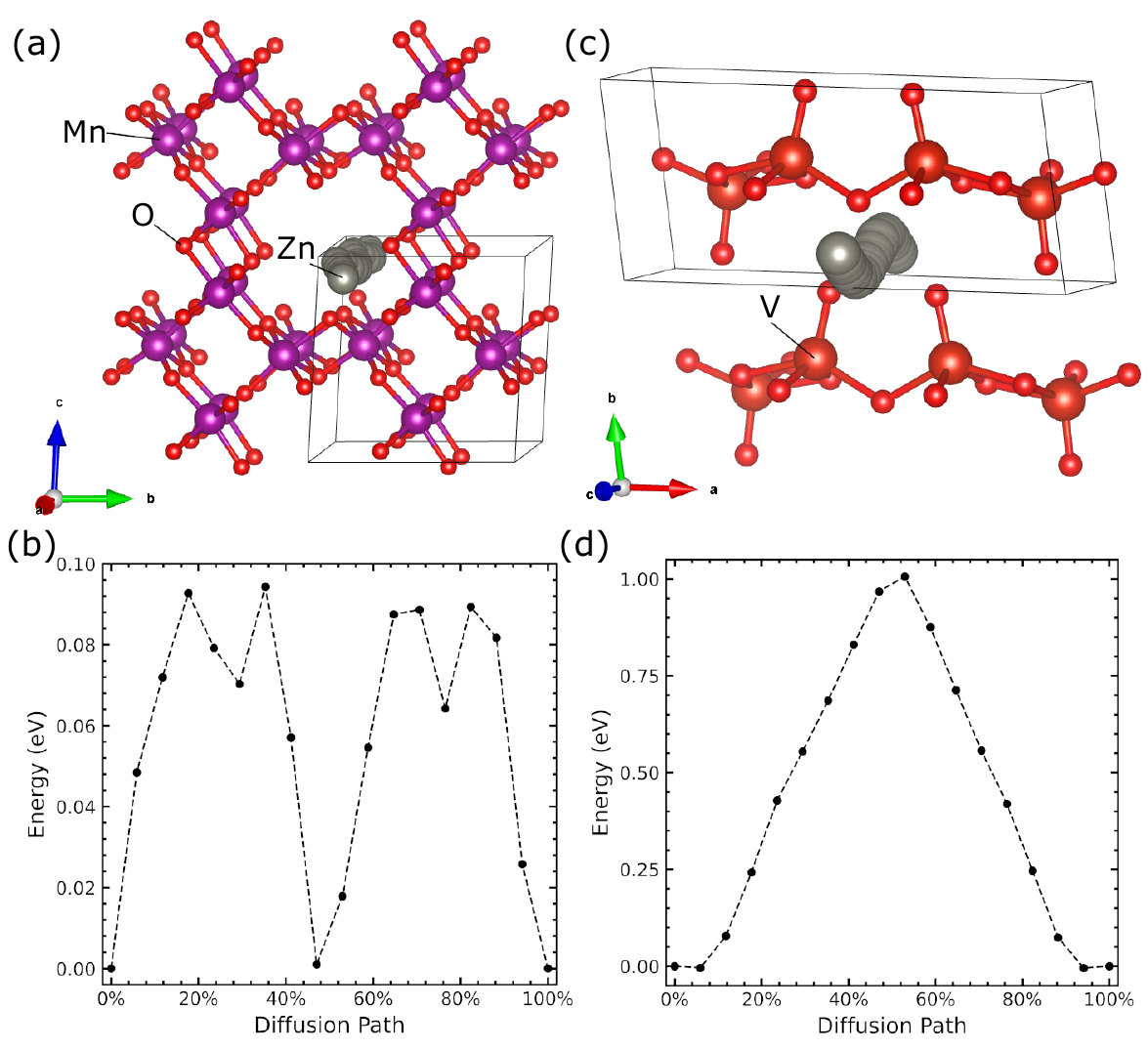}
    \caption{NEB lowest energy path for \ce{Zn^{2+}} diffusion in \ce{\alpha-MnO2} (a, b) and \ce{\alpha-V2O5} (c, d). The relative PBE+D3+$U$ total energy change along the path allows to evaluate the diffusion energy barrier.}
    \label{fig:NEB-MnO2-V2O5}
\end{figure}

In order to investigate how facile \ce{Zn^{2+}} transport can be expected during intercalation into the cathode material, its diffusion energy barrier in \ce{\alpha-V2O5} and \ce{\alpha-MnO2} was calculated. The converged path for \ce{Zn^{2+}} diffusion and its associated energy profile can be seen on Fig.~\ref{fig:NEB-MnO2-V2O5} for both materials. For these simulations supercells with only one \ce{Zn} atom were constructed, which resulted in \ce{ZnV4O10} and \ce{ZnMn8O16} stoichiometries. A full structure relaxation was performed for these compounds previous to the NEB calculation, so as to correctly capture its favourable intercalation ordering. The calculated energy barriers for \ce{Zn^{2+}} diffusion in \ce{\alpha-V2O5} and \ce{\alpha-MnO2} were of 1.0 and 0.1~eV, respectively. The large barrier height disparity can be linked to structural differences between the materials. \ce{\alpha-V2O5} is a layered system in which \ce{Zn} strongly interacts with both layers on its minimal energy path (Fig.~\ref{fig:NEB-MnO2-V2O5}c). In contrast, the characteristic $2 \times 2$ tunnels of \ce{\alpha-MnO2} have more space, and Zn atoms are partly bonded to one side of the tunnel (Fig.~\ref{fig:NEB-MnO2-V2O5}a).

A comparison can be made with other reported energy barriers for \ce{Zn^{2+}} diffusion within the systems analysed here. Prior first-principle studies reported much lower values of the barrier 0.305~eV \cite{Gautam_CC_51_2015} and 0.584~eV \cite{Sandagiripathira_PCCP_24_2022} for \ce{\alpha-V2O5} based on DFT-PBE calculations.  The discrepancy for \ce{\alpha-V2O5} can be partially explained by omission of the Hubbard $U$ correction in Refs.~\citenum{Gautam_CC_51_2015,Sandagiripathira_PCCP_24_2022}. We observed a decrease of the energy barrier by ca.~25\% when the $U$ correction is not applied. Another aspect is related to structural distortions of the host \ce{\alpha-V2O5} lattice in response to \ce{Zn} insertion. As can be seen on Fig.~\ref{fig:struct-benchmark}(c) for \ce{ZnV2O5} and Fig.~\ref{fig:NEB-MnO2-V2O5}(c) for \ce{ZnV4O10}, the \ce{[VO5]} pyramids get titled upon \ce{Zn} insertion in attempt to establish \ce{Zn-O} bonds. This phenomenon is well documented for intercalation of other ionic species (e.g.,  \ce{Li+} and \ce{Mg^{2+}}  \cite{Horrocks_JMCA_1_2013,Gautam_CM_27_2015}) in \ce{\alpha-V2O5}. These structural distortions impede a direct linear diffusion path between the layers, thus resulting in a curved \textit{S}-shaped path with a larger barrier. Graphical images presented in Refs.~\citenum{Gautam_CC_51_2015,Sandagiripathira_PCCP_24_2022} suggest that such structural distortions due to \ce{Zn} presence could have been omitted in prior works, while the lack of calculated structures therein precludes us from reproducing data (see Sec.~``Supporting Information Available" for data availability pertaining to this work). In the case of \ce{\alpha-MnO2}, \citet{Putro_EA_345_2020} reported the barrier of 0.34~eV for \ce{Zn} migration evaluated on a path that follows through the centre of the characteristic $2 \times 2$ void rather than selecting energetically more preferable coordination of the add-atom.

\citet{Rong_CM_27_2015} arrived to 0.525~eV as a reference maximum value for an intercalated ion energy barrier by analyzing different aspects pertaining to \ce{Li}-ion batteries, for example its (dis)charge rate and active material size, when considering the ion diffusion as a random walk. From this we would expect efficient \ce{Zn} intercalation in \ce{\alpha-MnO2} only. As the energy barrier for \ce{\alpha-V2O5} is an order of magnitude higher, we can anticipate the \ce{Zn^{2+}} diffusion in \ce{\alpha-V2O5} to be drastically different ($10^{15}$ fold slower at the room temperature). However, numerous experiments \cite{Chen_AMI_6_2019,Selvakumaran_JMCA_7_2019, Yi_CCR_446_2021,Blanc_J_4_2020} evidence that both materials demonstrate similar performance as cathodes for RAZIB. The discordance between the barrier height and performance of \ce{\alpha-MnO2} and \ce{\alpha-V2O5} cathode materials calls for dissolution and subsequent re-deposition of the cathode material in the aqueous electrolyte as possible alternative energy storage mechanism (following Refs.~\citenum{Selvakumaran_JMCA_7_2019,Blanc_J_4_2020}) that is not sensitive to Zn diffusion. Eventually, we are unable to conclude on the barrier height as a useful design metric for RAZIB cathode materials.

\subsection{Role of hydration in incorporation of Zn}\label{sec:results:role-hydration}

\begin{figure}
    \includegraphics{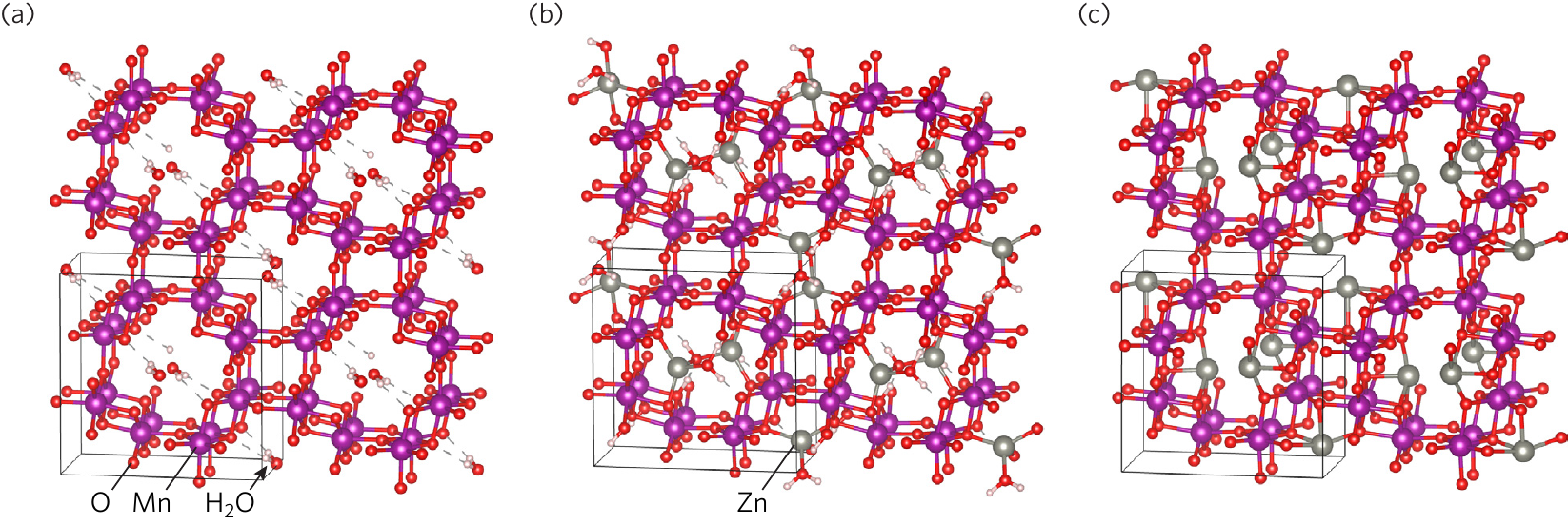}
    \caption{Structures of (a) \ce{$\alpha$-Mn4O8.H2O}, (b) \ce{ZnMn4O8.H2O}, and (c) \ce{ZnMn4O8}.}
    \label{fig:ZnMn4O8-hydr}
\end{figure}

Here we explore the role of hydration in incorporation of \ce{Zn^{2+}} ions in \ce{\alpha-MnO2} with the hope to improve the agreement with experiment (Table~\ref{tab:Zn-potential-benchmark}). Calculations of \ce{Zn^{2+}} deintercalation potential where performed for structures shown in Fig.~\ref{fig:ZnMn4O8-hydr}. Those structures represent stoichiometry of a fully charged (Fig.~\ref{fig:ZnMn4O8-hydr}a), a half discharged  hydrated (Fig.~\ref{fig:ZnMn4O8-hydr}b) and anhydrous (Fig.~\ref{fig:ZnMn4O8-hydr}c) states. The partly discharged state was used as it was not possible to accommodate more Zn ions and water molecules at the same time.

By comparing the total energies of \ce{$\alpha$-Mn4O8.H2O} with the anhydrous \ce{\alpha-MnO2} and solid water, we conclude that the hydration enthalpy is ca. 0.4~eV per \ce{H2O} molecule. The positive enthalpy suggests that incorporation of water is energetically unfavourable. Similar comparison of total energies for structures with zinc \ce{ZnMn4O8.H2O} vs \ce{ZnMn4O8} and solid water yields the hydration enthalpy of 0.13~eV per \ce{H2O} molecule. This trend indicates that the hydration becomes more favourable when accompanied by intercalation of Zn.

Using total energies of the structures in Fig.~\ref{fig:ZnMn4O8-hydr}a,b for the charged and discharged phases, respectively, we obtained the \ce{Zn^{2+}} deintercalation potential of 1.0~V vs \ce{Zn^{0}/Zn^{2+}} at the PBE+D3+$U$ level of theory. As a reference, we also evaluated the \ce{Zn^{2+}} deintercalation potential for the very similar anhydrous structure (Fig.~\ref{fig:ZnMn4O8-hydr}c) as 0.9~V vs \ce{Zn^{0}/Zn^{2+}}. Even though the coordination of Zn ions is improved in the hydrated structure, the hydration does not improve the agreement with experiment for the \ce{Zn^{2+}} deintercalation potential (Table~\ref{tab:Zn-potential-benchmark}), contrary to calculations carried out by \citet{Le_AAMI_13_2021}. It should be noted that Ref.~\citenum{Le_AAMI_13_2021} considered a higher number of incorporated water molecules, which resulted in breaking the characteristic \ce{\alpha-MnO2} $2 \times 2$ tunnels. This could explain their ability to achieve a greater \ce{Zn^{2+}} deintercalation potential. However, to the best of our knowledge, the literature lacks reports supporting either of those $\alpha$-like \ce{Zn_{$x$}MnO2.$y$H2O} structures derived directly from experimental crystallographic techniques. Hence, our analysis of hydrated structures was limited to those where the structural features of \ce{\alpha-MnO2}  were preserved.

\subsection{Selection of alternative cathode materials}\label{subsec:materials selection}

\begin{figure}
    \centering
    \includegraphics{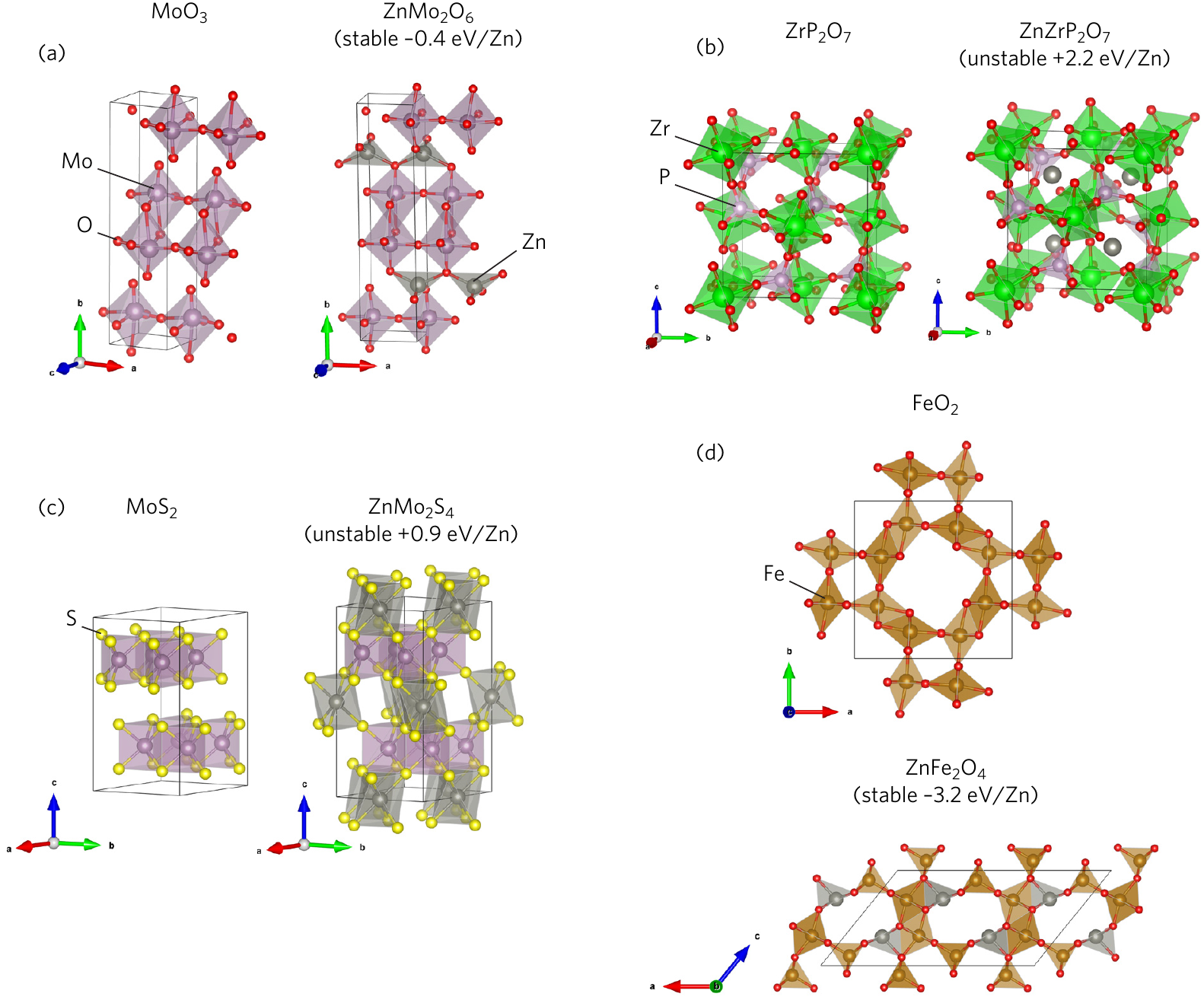}
    \caption{Structures of alternative cathode materials without and with intercalated Zn: (a) \ce{MoO3} and \ce{ZnMo2O6}, (b) \ce{ZrP2O7} and  \ce{ZnZrP2O7}, (c) \ce{MoS2} and \ce{ZnMo2S4}, (d) \ce{$\alpha$-FeO2} and $\alpha$-like \ce{ZnFe2O4}. Feasibility of Zn intercalation is expressed as energy per Zn atom calculated using Eq.~\eqref{eq:intercalation energy} at PBE+D3+$U$ level.}
    \label{fig:struct-explored}
\end{figure}

We selected two layered materials (\ce{MoS2} and \ce{MoO3}) and two materials with tunnels or voids (\ce{ZrP2O7} and \ce{$\alpha$-FeO2}) as shown in Fig.~\ref{fig:struct-explored}. This selection represents TM oxides and sulfides with a crystal structures suitable to host Zn (the atomic diameter of Zn is 2.7~{\AA} \cite{Slater_JCP_41_1964}). The initial crystal structures were obtained from Springer Materials \cite{Villars_SpringerMaterials_2016} or the crystallography open database \cite{Downs_AM_88_2003,Grazulis_JAC_42_2009,Grazulis_NAR_40_2011}. \ce{ZrP2O7} (Fig.~\ref{fig:struct-explored}b) and \ce{$\alpha$-FeO2} (Fig.~\ref{fig:struct-explored}d) have voids of the size 3.1 and 4.1~{\AA}, respectively. For comparison, \ce{\alpha-MnO2} has the tunnel size of 3.9~{\AA} based on \citet{Slater_JCP_41_1964} atomic radii.

Alternatively, the intercalation of \ce{Zn^{2+}} can take place via intercalation of the hydrated ion \cite{Lee_CC_51_2015,Kundu_NE_1_2016,Shin_AEM_9_2019,Adams_patent_layered_2020,Hou_JMC_8_2020}. \citet{Cauet_JCP_132_2010} concluded that the first hydration shell involves six water molecules arranged in an octahedral geometry \ce{Zn^{2+} + 6H2O} (see Fig.~7 therein). Therefore, we can estimate diameter of the first hydration shell as 6.7~{\AA} based on the \ce{Zn^{2+} - O_{I}} distance of 2.1~{\AA} \cite{Cauet_JCP_132_2010}, the \ce{O_{I} - H} distance of 1~{\AA}, and the hydrogen radius of 0.25~{\AA} \cite{Slater_JCP_41_1964}. Given the size of the hydration shell, only layered structures have enough spacing to accommodate fully hydrated \ce{Zn^{2+}} ions.

The choice of \ce{$\alpha$-FeO2} was inspired by \citet{Brady_JPCC_123_2019} who explored possibilities of TM substitutions in \ce{\alpha-MnO2}.  There are two minerals Akaganeite and Schwertmannite with the \ce{$\alpha$-FeO2} tunnel structure \cite{Post_AM_88_2003,FernandezMartinez_AM_95_2010}, however both are minerals of Fe(III); Fe(IV)-oxides do not occur in nature. The spinel phase of \ce{ZnFe2O4} exists and showed some performance in Li-ion batteries \cite{Zhang_CM_29_2017}. However, its narrow tunnels (2.1~{\AA}) are not ideal for reversible transport of \ce{Zn^{2+}} ions.

\subsection{Zn intercalation affinity}\label{subsec:Zn intercalation affinity}

There are no experimental structures of 
\ce{MoS2}, \ce{MoO3}, \ce{ZrP2O7}, and \ce{FeO2} with Zn that would closely resemble the pristine structure. Zn intercalation sites were chosen after inspecting voids in the host structures. When placing Zn into the structure we tried to match its natural coordination (four-, five-, and six-fold) when possible. Subsequently, all structures were fully relaxed to minimize forces and stresses.

Structures of cathode materials with and without Zn are shown in Fig.~\ref{fig:struct-explored}. Upon Zn insertion all structures retained their similarity to the host material with the exception of \ce{$\alpha$-ZnFe2O4} (Fig.~\ref{fig:struct-explored}d). We observed this behaviour also in \ce{$\alpha$-ZnMn2O4} (Fig.~\ref{fig:struct-benchmark}b) and attributed it to a transformation from the structure with tunnels to a layered structure similar to \ce{$\alpha$-K$_x$MnO2} structures \cite{Jiao_JMCA_8_2020}.

The Zn intercalation energy $(\Delta H_{\text{Zn}})$ calculated according to Eq.~\eqref{eq:intercalation energy} is given alongside of each structure in Fig.~\ref{fig:struct-explored}. After evaluating the Zn intercalation energy, we conclude that \ce{MoS2}/\ce{ZnMo2S4} and \ce{ZrP2O7}/\ce{ZnZrP2O7} are not suitable for hosting Zn. This result corroborates failed experimental attempts to observe Zn intercalation and energy storage in unmodified \ce{MoS2} \cite{Liu_CC_53_2017}. However, there are other reports of successful Zn intercalation in a modified \ce{MoS2} with defects (transition-metal vacancies) \cite{Lee_SM_5_2020,Li_AN_13_2019} or a metastable (metallic) phase 1T' \cite{Liu_EA_410_2022}.  In the case of defects Zn passivates them. For example, two \ce{Zn^{2+}} ions absorbed per one \ce{Mo^{4+}} vacancy, which is possible in layered structures. However, the defect concentration should be very high to reach a significant capacity, which raises concerns about the structural integrity of the crystal. We continue with \ce{FeO2}/\ce{ZnFe2O4} and \ce{MoO3}/\ce{ZnMo2O6} as potential candidates.

\subsection{Volume change upon Zn intercalation}\label{subsec:Volume change}

Any volumetric changes within the cathode material provide insight into the cyclic stability and longevity of the battery. Large distortions and volume changes during charge or discharge can lead to uneven strain within the cathode material causing poor mechanical stability, cracking and capacity fading \cite{Liu_AN_6_2012}. In Li-ion commercial batteries the volume change of the cathode material does not exceed 8\% upon lithiation \cite{Biasi_JPS_362_2017}. We will take this value as a guideline in our study, however we treat deviations as a warning rather than a requirement for materials selection.

The unit cell of \ce{\alpha-MnO2} and \ce{FeO2} undergoes drastic changes upon Zn intercalation (Figs.~\ref{fig:struct-benchmark}b and \ref{fig:struct-explored}d), even though the overall volume change is only 9\% (Table~\ref{tab:characteristics summary}). This renders both materials as prone to mechanical instabilities along with \ce{\alpha-V2O5} which has a large volume change. Among all cathode materials in this study only \ce{MoO3} falls within the acceptable volume change (under 8\%) during Zn intercalation-deintercalation.

\begin{table}
\caption{Summary of characteristics for cathode materials. All computational values correspond to PBE+D3+$U$ level of theory. Problematic entries are highlighted in bold.}
\label{tab:characteristics summary}
\begin{tabular}{ >{\raggedright}p{50mm} c c c c c c }
    \hline
    Characteristic & \ce{\alpha-MnO2} & \ce{\alpha-V2O5} & \ce{MoS2} & \ce{ZrP2O7} & \ce{MoO3} & \ce{FeO2} \\
    \hline
    Structural feature \vspace{10pt} & tunnels $2 \times 2$ & layered & layered & cavity & layered & tunnels $2 \times 2$  \\
    Thermodynamic stability without Zn (eV/atom above hull) \vspace{10pt} & 0 & 0 & 0 & 0 & 0 & \textbf{0.25}  \\
    Feasible to synthesise without Zn \vspace{10pt} & yes & yes & yes & yes & yes & \textbf{no}  \\
    Zn insertion per formula unit (atom) \vspace{10pt} & 0.5 & 1 & 0.5 & 1 & 0.5 & 0.5\\
    Energy storage capacity (mA~h~g$^{-1}$) \vspace{10pt} & 310 & 300 & 170 & 200 & 190 & 310 \\
    Redox couple \vspace{10pt} & \ce{Mn^{4+/3+}} & \ce{V^{5+/4+}} & \ce{Mo^{4+/3+}} & \ce{Zr^{4+/2+}} & \ce{Mo^{6+/5+}} & \ce{Fe^{4+/3+}} \\
    Zn intercalation energy (eV/Zn) \vspace{10pt} & $-2.2$  & $-2.0$ & \textbf{+0.9} & \textbf{+2.2} & $-1.4$ & $-3.2$\\
    Thermodynamic stability with Zn (eV/atom above hull) \vspace{10pt} & \textbf{0.21} & 0.07 & --- & --- & 0.14 & \textbf{0.18} \\
    Calculated charge potential (V vs \ce{Zn^{0}/Zn^{2+}}) \vspace{10pt} & $1.1^{+0.4}$\footnote{The correction of $\sim$0.4~V is added to account for a nonoptimal coordination of Zn atoms in a model that represents the discharged structure.} & 1.0 & --- & --- & 0.7 & $\mathbf{1.6^{+0.4\text{a}}}$ \\
    Volume change upon Zn insertion (ln \%) \vspace{10pt} & \textbf{9} & \textbf{15} & --- & --- & 4 & \textbf{9} \\
    Electrochemical stability: corrosion/dominance of another solid phase (\% of area in Pourbaix diagram) \vspace{10pt} & \textbf{60}/0 & \textbf{10}/\textbf{85} & --- & --- & \textbf{80}/0 & 0/\textbf{100} \\
    Charge potential is within the cathode electrochemical stability region \vspace{10pt} & yes & \textbf{no} & --- & --- & \textbf{no} & \textbf{no} \\
    Diffusion barrier (eV) \vspace{10pt} & 0.1 & 1.3 & --- & --- & --- & --- \\
    \hline
\end{tabular}
\end{table}

\subsection{\ce{Zn^{2+}} deintercalation potentials}\label{subsec:Deintercalation potentials}

We computed the average charge (or deintercalation) potential according to Eq.~\eqref{eq:deintercalation potential} for cathode materials that showed a favorable Zn intercalation. The deintercalation potential translates into the open circuit voltage measured against the zinc-metal anode (\ce{Zn^{0}}/\ce{Zn^{2+}}). 

According to Eq.~\eqref{eq:deintercalation potential} the potential is 1.6 and 0.7~V vs \ce{Zn^{0}/Zn^{2+}} for \ce{$\alpha$-FeO2}/\ce{ZnFe2O4} and \ce{MoO3}/\ce{ZnMo2O6}, respectively, at the PBE+D3+$U$ level of theory. Cyclic voltammetry of \ce{MoO3} in a \ce{ZnSO4} aqueous electrolyte shows oxidation and reduction peaks between 0.5 and 0.9~V vs \ce{Zn^{0}/Zn^{2+}} \cite{Liu_CC_53_2017}.

We note increase in the theoretical charge potential from 1.1 (see Table~\ref{tab:characteristics summary}) to 1.6~V vs \ce{Zn^{0}/Zn^{2+}} when substituting \ce{Mn^{4+/3+}} with \ce{Fe^{4+/3+}} in the hollandite-like structure. This trend is consistent with ordering of redox energies of 3d TM ion couples (see Fig.~1 in Ref.~\citenum{MoorheadRosenberg_IC_52_2013}). Since PBE+D3+$U$ underestimated the experimental charge potential in \ce{\alpha-MnO2}/\ce{ZnMn2O4} by 0.4~V (see Table~\ref{tab:Zn-potential-benchmark}), we can expect a higher experimental charge potential (ca. 2~V vs \ce{Zn^{0}/Zn^{2+}}) for the \ce{$\alpha$-FeO2}/\ce{ZnFe2O4} cathode material.

Considering the use of a mildly acidic aqueous electrolyte, the desired voltage range is from 1 to 1.8~V vs \ce{Zn^{0}/Zn^{2+}}. The upper bound is governed by the oxygen evolution potential. Thus, oxides with the \ce{Fe^{4+/3+}} redox couple, in general, may not be suitable for Zn-ion rechargeable batteries with the aqueous electrolyte.

Calculations reported in this paper are performed without taking the aqueous medium into account explicitly. \citet{Shin_AEM_9_2019} reported that the computed voltages were higher in anhydrous \ce{V6O_{13}} and the voltages of hydrated structure overlapped with experimental results, while the voltages computed by \citet{Wu_JMCA_7_2019} show the opposite trend of \ce{\alpha-V2O5} resulting in overestimation of the voltages in hydrated structures. However, intercalation cathodes have been modelled accurately without including the solvation sphere around the cation \cite{Jiao_JMCA_8_2020}. Our own attempts reported in Sec.~\ref{sec:results:role-hydration} did not improve calculations of the charge potential. Hence, the solvation sphere is not included in the computations.

We can conclude that \ce{MnO2}, \ce{\alpha-V2O5}, and \ce{MoO3} have favourable charge potentials (Table~\ref{tab:characteristics summary}).

\subsection{Thermodynamic stability of structures before and after intercalation}\label{subsec:Thermodynamic stability}

Ideally, both the pristine cathode material and its Zn intercalated counterpart should show thermodynamic stability. So far, we used Eq.~\eqref{eq:intercalation energy} to assess the Zn affinity. The negative value of the intercalation energy $\Delta H_\text{Zn}$ is a necessary condition for thermodynamic stability of the intercalated compound, but it is not a sufficient condition. To prove the thermodynamic stability we need to show that the intercalated compound is located on the energy convex hull. As of now, even previously successful cathode materials (\ce{\alpha-MnO2} and \ce{\alpha-V2O5}) were unable to meet this requirement for their Zn intercalated derivatives being 0.05$-$0.2~eV/atom above the hull (Fig.~\ref{fig:deltaH-MnO2-V2O5}c,d) in DFT calculations at the PBE+D3+$U$ level of theory.

\begin{figure}
	\includegraphics{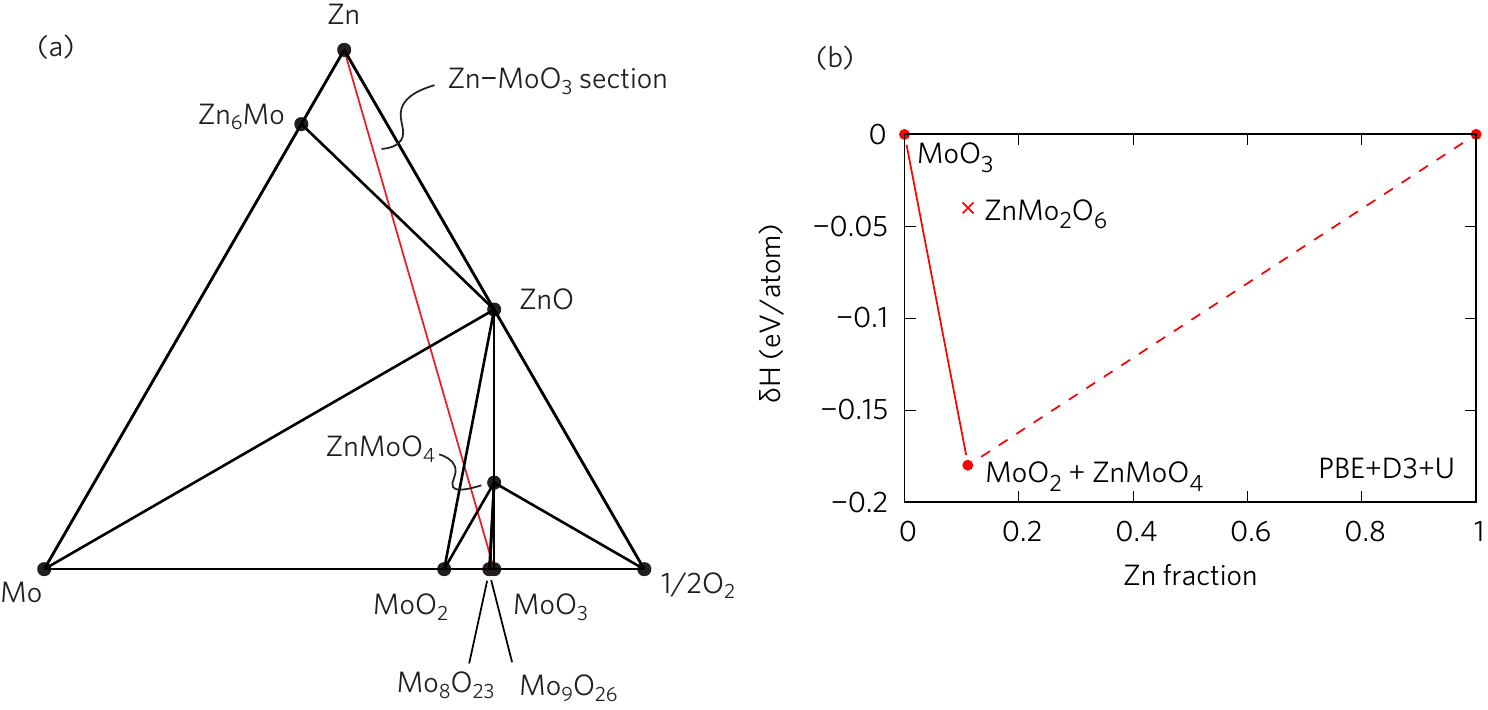}
	\caption{Phase diagrams: (a) Ternary \ce{Zn-Mo-O} phase diagram reproduced from Materials project \cite{Ong_CM_20_2008,Jain_PRB_84_2011}. (b) \ce{Zn-MoO3} section of the phase diagram calculated in this work at the PBE+D3+$U$ level.}
    \label{fig:deltaH-MoO3}
\end{figure}

Figure~\ref{fig:deltaH-MoO3}a shows a ternary \ce{Zn-Mo-O} phase diagram from the Materials Project \cite{Ong_CM_20_2008,Jain_PRB_84_2011} obtained by a mixed PBE/PBE+$U$ scheme (including empirical corrections). The diagram correctly predicts major phases, such as \ce{MoO2}, , \ce{Mo8O23}, \ce{MoO3}, and \ce{ZnMoO4} known experimentally \cite{Predel_Mo-O_phase-diagram_1997,Reichelt_ZAAC_626_2000,Soehnel_ZAAC_623_1997}, while failed to capture \ce{Mo9O24}, \ce{Mo4O11},  \ce{Zn2Mo3O8}, \ce{Zn3Mo2O9}. There is no stable phase with \ce{ZnMo2O6} stoichiometry which would represent the discharged state of the cathode material. According to the phase diagram (Fig.~\ref{fig:deltaH-MoO3}b) it would decompose into \ce{MoO2 + ZnMoO4} which places \ce{ZnMo2O6} into unstable position ca.~0.14~eV/atom above the convex hull (Fig.~\ref{fig:deltaH-MoO3}b).

\begin{figure}
	\includegraphics{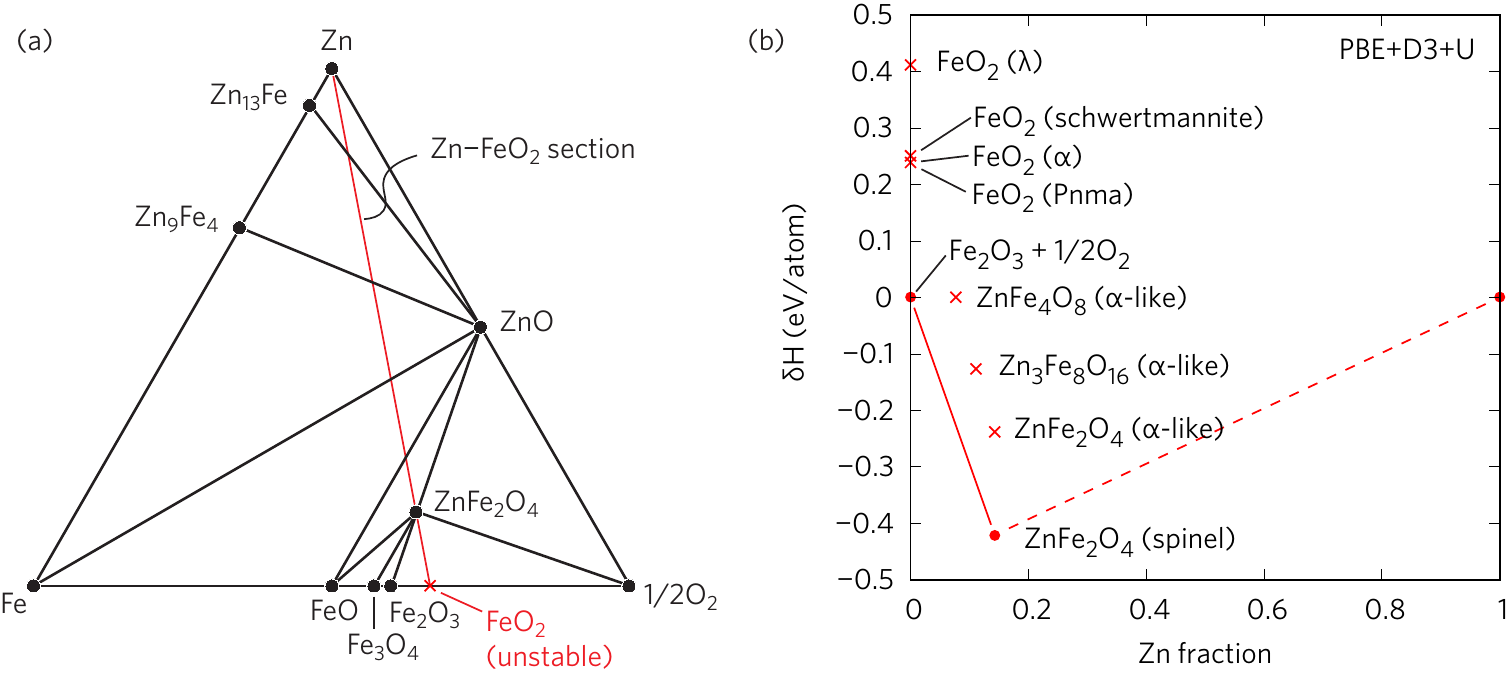}
	\caption{Phase diagrams: (a) Ternary \ce{Zn-Fe-O} phase diagram reproduced from Materials project \cite{Ong_CM_20_2008,Jain_PRB_84_2011}. (b) \ce{Zn-FeO2} section of the phase diagram with \ce{Zn$_{x}$FeO2} phases calculated in this work at the PBE+D3+$U$ level.}
    \label{fig:deltaH-FeO2}
\end{figure}

Figure~\ref{fig:deltaH-FeO2}a shows a ternary \ce{Zn-Fe-O} phase diagram from the Materials Project \cite{Ong_CM_20_2008,Jain_PRB_84_2011}. The diagram correctly predicts phase \ce{Fe3O4} and \ce{Fe2O3} stable at ambient conditions, but also shows a high-temperature phase FeO \cite{Kubaschewski_Fe-O_phase-diagram_2007}. The Materials Project data and also our own calculations in Fig.~\ref{fig:deltaH-FeO2}b preclude existence of the \ce{FeO2} phase at ambient conditions (0.25~eV/atom above the hull), which correlates with experimental results \cite{Hu_N_534_2016}. The intercalated $\alpha$-like \ce{ZnFe2O4} structure is also above the hull by 0.18~eV/atom, which applies to structures with an intermediate Zn content too. Thus, the \ce{Zn/FeO2} is the only couple where both intercalated and deintercalated structures are unstable.

It is instructive to put calculated energies relative to the convex hull in the context of existing battery materials. In the case of \ce{Li-Co-O} ternary system, the layered \ce{LiCoO2} phase is thermodynamically stable \cite{MSIEureka2016:sm_msi_r_10_029498_01}, while the deintercalated \ce{CoO2} structure is not. The \ce{CoO2} solid is 0.16~eV/atom above the energy convex hull based on assessment of CALPHAD enthalpies presented in Table~3.2 of Ref.~\citenum{Chang_PhD_thesis_2013}. Thus we can use this value as a threshold when the thermodynamic stability should raise a concern (bold values in Table~\ref{tab:characteristics summary}). We can conclude that only \ce{\alpha-V2O5} and \ce{MoO3} meet expectations on their thermodynamic stability in charged and discharged states (Table~\ref{tab:characteristics summary}).

\subsection{Electrochemical stability}

Understanding the electrochemical stability of cathode structures in an aqueous environment during charge and discharge operation is critical to designing new electrode materials with extended operational lifetimes. The equilibrium electrochemical stability of a \ce{metal-H2O} system can be inferred from Pourbaix diagrams. For mildly acidic zinc-sulphate electrolyte, the region of interest is pH~3$-$5 \cite{Bischoff_JES_167_2020} and potential $E_{\pm} \approx E\pm 0.5$~V, where $E$ is the average charge potential given by Eq.~\eqref{eq:deintercalation potential}. The aqueous concentration of TM species of 0.1~M is assumed based on typical concentrations of \ce{MnSO4} used in experimental \ce{Zn/MnO2} battery research \cite{Pan_NE_1_2016,Chamoun_ESM_15_2018,Bischoff_JES_167_2020}. Construction of fully \textit{ab initio} Pourbaix diagrams is still out of reach due to difficulties with correlation effects on TM and finite-temperature thermodynamic properties \cite{Jain_PRB_84_2011,Persson_PRB_85_2012,Zeng_JPCC_119_2015,Wang_nCM_6_2020}. Therefore, we will rely on our analysis of experimental Pourbaix diagrams.

\begin{figure}
	\includegraphics{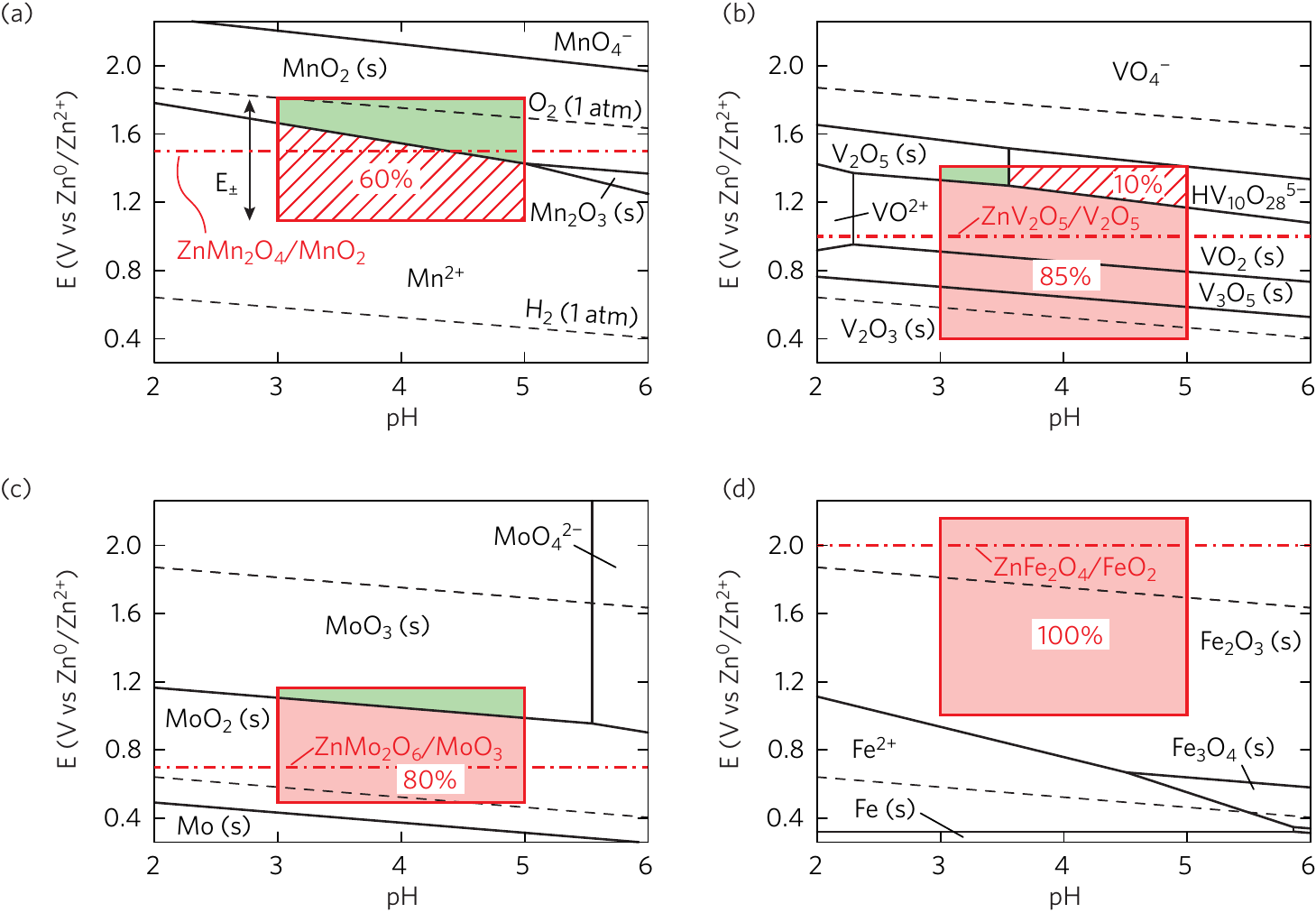}
	\caption{Pourbaix diagrams of (a) \ce{Mn-H2O}, (b) \ce{V-H2O}, (c) \ce{Mo-H2O}, and (d) \ce{Fe-H2O} from FactSage thermochemical software \cite{Bale_C_54_2016}. Concentration of aqueous species is 0.1~M. The red square shows the range of pH and potential during an operating charge/discharge cycle. Within this range the green colour shows boundaries of stability of the pristine (Zn free) cathode material. The pink and hatched areas highlight problematic regions of corrosion and areas where another solid phase (different from the cathode material) is predominant. The percent values indicate a relative share of each area. The red dash-dotted line shows the \ce{Zn^{2+}} deintercalation potential from Table~\ref{tab:characteristics summary}.}
    \label{fig:Pourbaix}
\end{figure}

The electrochemical stability of \ce{MnO2} in the context of RAZIBs was recently analysed \cite{Rubel_JPCC_126_2022} and the charge/discharge cycle was mapped on the \ce{Mn-H2O} Pourbaix diagram \cite{Bischoff_JES_167_2020}. \ce{MnO2} is stable at charging potentials, but enters the corrosion region during discharge (Fig.~\ref{fig:Pourbaix}a) which limits its stability within the relevant pH/$E_{\pm}$ window. It is even claimed that some of \ce{Zn/MnO2} capacity can originate from a reversible dissolution of the cathode material \cite{Lee_SR_4_2014,Wu_EES_13_2020,Moon_AS_8_2021,Rubel_JPCC_126_2022}.

\ce{\alpha-V2O5} stability exists within a narrow range of both pH and potentials (Fig.~\ref{fig:Pourbaix}b). Its stability boundaries are extremely sensitive to the concentration of aqueous V species. At the concentration of V aqueous species of 0.1~M, \ce{\alpha-V2O5} is unstable above pH~3$-$4 and is prone to corrosion. \ce{\alpha-V2O5} also becomes unstable at lower potentials $E<1.2\!-\!1.3$~V vs \ce{Zn^{0}/Zn^{2+}} in favour of \ce{VO2} (Fig.~\ref{fig:Pourbaix}b). The predominance of another solid phase can make the process of dissolution and subsequent re-deposition of the cathode material not fully reversible thereby contributing to capacity fade. Although, \ce{Zn/V2O5} rechargeable aqueous batteries have a good capacity retention at high current (5~A~g$^{-1}$) \cite{Zhang_AEL_3_2018}, the capacity degrades very fast at lower currents (0.1$-$0.2~A~g$^{-1}$) \cite{Yang_JMCA_8_2020} due to the cathode dissolution as one of the main factors. Claims of experimental gravimetric energy density (ca.~400~mA~h~g$^{-1}$) \cite{Zhang_AEL_3_2018} in the excess of the theoretical limit (300~mA~h~g$^{-1}$, Table~\ref{tab:characteristics summary}) are not uncommon for  \ce{Zn/V2O5} suggesting a possible contribution of a reversible cathode dissolution to the capacity similar to \ce{MnO2} \cite{Lee_SR_4_2014,Moon_AS_8_2021,Rubel_JPCC_126_2022}.

\ce{MoO3} is stable against corrosion up to pH~5.5 (Fig.~\ref{fig:Pourbaix}c). However, \ce{MoO2} takes over the \ce{MoO3} phase at potentials below 0.9$-$1.0~V vs \ce{Zn^{0}/Zn^{2+}}. The predicted charge potential for \ce{Zn/MoO3} is 0.7~V vs \ce{Zn^{0}/Zn^{2+}}. Thus, \ce{MoO3} is not stable within the full relevant range of parameters. Experimentally a rapid degradation of \ce{Zn/MoO3} deliverable capacity was observed after the first 1$-$2 cycles \cite{Liu_CC_53_2017}.

\ce{FeO2} is not present on the Pourbaix diagram (Fig.~\ref{fig:Pourbaix}d), which is expected since \ce{FeO2} is unstable. \ce{Fe2O3} is the only stable solid phase present within the battery operating window, with the discharge potential staying above the corrosion region.

Another important consideration for deployment as cathodes in aqueous zinc ion batteries is the electrochemical stability of aqueous electrolyte against decomposition. It should be noted that the discharge potential for \ce{Zn/V2O5} and \ce{Zn/MoO3} is bound by the hydrogen evolution line (Fig.~\ref{fig:Pourbaix}b,c). The charge potential for \ce{Zn/MnO2} is at the oxygen evolution line, while it dangerously exceeds the oxygen evolution line for \ce{Zn/FeO2} (Fig.~\ref{fig:Pourbaix}a,d).

Furthermore, the region of electrochemical stability for \ce{\alpha-V2O5}, \ce{MoO3} and \ce{FeO2} solids does not overlap with the calculated charge potentials (see Fig.~\ref{fig:Pourbaix}, dash-dotted lines). Therefore, it is possible to infer that Zn intercalation into the charged cathode will not be a direct and unopposed reaction as it is desired for battery application. Instead, other solid or aqueous phases will compete for occurring within the operating range of potentials/pH. This shortcoming can lead to a low cycling capability and/or capacity fade due to irreversible cathode decomposition.

Thus, none of the materials analysed fully meet expectations on their electrochemical stability (Table~\ref{tab:characteristics summary}), indicating the present challenges in the discovery of suitable host materials. It also highlights the necessity of combining theoretical and experimental efforts for exploring broader material classes in the search for novel RAZIB cathodes. 

\section{\label{sec:level1}Conclusion}

We proposed, verified, and applied a comprehensive set of indicators to screen candidate cathode materials for Zn-ion aqueous rechargeable batteries. The indicators capture feasibility of \ce{Zn^{2+}} intercalation and transport within the material, the thermodynamic stability of charged and discharged material, electrochemical stability of the cathode material and electrolyte, volume change, and energy storage capacity. Most characteristics can be computed from first principles with the exception of Pourbaix diagrams for the electrochemical stability that currently require experimental inputs. The approach was verified using \ce{\alpha-MnO2} and \ce{\alpha-V2O5} as benchmark, the two well studies cathode materials. The screening protocol was applied to \ce{MoS2}, \ce{ZrP2O7}, \ce{MoO3}, \ce{FeO2} that covered layered, tunnel, and hollow structures to identify alternative cathode materials for Zn-ion aqueous rechargeable batteries. Neither \ce{MoS2} nor \ce{ZrP2O7} were able to host Zn (had a positive Zn intercalation energy). Other materials (\ce{MoO3} and \ce{FeO2}) were able to host Zn, but had other shortcomings (thermodynamic instabilities, charge overpotential leading to electrolyte instabilities, electrochemical instabilities). Electrochemical stability was the most stringent criterion that none of the explored materials were able to meet fully. The second and third most demanding requirements were (i) less than 8\% volume change upon Zn intercalation and (ii) thermodynamic stability of the Zn intercalated structure (not more than 0.16~eV/atom above the energy convex hull). The calculated energy barriers for \ce{Zn^{2+}} diffusion in \ce{\alpha-V2O5} and \ce{\alpha-MnO2} show a large disparity despite both materials demonstrating similar performance as cathodes in RAZIBs. Thus, we are unable to endorse the barrier height as a useful design metric, which is possibly related to competing electrochemical energy storage mechanisms not sensitive to Zn diffusion (e.g., dissolution and subsequent re-deposition of the cathode material in the aqueous electrolyte). While the materials explored in this work did not emerge as promising for application in RAZIBs from a materials property standpoint, the tools and techniques developed in this work can be  applied more broadly to screen a wider array of potential material compositions and structures, with the goal of identifying next generation RAZIB cathode materials that can overcome performance and stability shortcoming plaguing this technology for grid scale energy storage applications.

\begin{acknowledgement}

O.R. would like to thank Dmitri V.~Malakhov (McMaster University) for stimulating discussions regarding electrochemical stability of materials. This work was supported by the Salient Energy, the NSERC Alliance program, and the Mitacs Globalink program. Calculations were performed using the Compute Canada infrastructure supported by the Canada Foundation for Innovation under John R. Evans Leaders Fund.

\end{acknowledgement}

\begin{suppinfo}

The raw data (VASP input and structure files) required to reproduce these findings are available in the form Zenado file repository \cite{Zenodo_10.5281/zenodo.7796317}.

\end{suppinfo}

\bibliography{bibliography}

\providecommand{\latin}[1]{#1}
\makeatletter
\providecommand{\doi}
  {\begingroup\let\do\@makeother\dospecials
  \catcode`\{=1 \catcode`\}=2 \doi@aux}
\providecommand{\doi@aux}[1]{\endgroup\texttt{#1}}
\makeatother
\providecommand*\mcitethebibliography{\thebibliography}
\csname @ifundefined\endcsname{endmcitethebibliography}
  {\let\endmcitethebibliography\endthebibliography}{}
\begin{mcitethebibliography}{106}
\providecommand*\natexlab[1]{#1}
\providecommand*\mciteSetBstSublistMode[1]{}
\providecommand*\mciteSetBstMaxWidthForm[2]{}
\providecommand*\mciteBstWouldAddEndPuncttrue
  {\def\EndOfBibitem{\unskip.}}
\providecommand*\mciteBstWouldAddEndPunctfalse
  {\let\EndOfBibitem\relax}
\providecommand*\mciteSetBstMidEndSepPunct[3]{}
\providecommand*\mciteSetBstSublistLabelBeginEnd[3]{}
\providecommand*\EndOfBibitem{}
\mciteSetBstSublistMode{f}
\mciteSetBstMaxWidthForm{subitem}{(\alph{mcitesubitemcount})}
\mciteSetBstSublistLabelBeginEnd
  {\mcitemaxwidthsubitemform\space}
  {\relax}
  {\relax}

\bibitem[Cabana \latin{et~al.}(2010)Cabana, Monconduit, Larcher, and
  Palac{\'{\i}}n]{Cabana_AM_22_2010}
Cabana,~J.; Monconduit,~L.; Larcher,~D.; Palac{\'{\i}}n,~M.~R. Beyond
  intercalation-based {Li}-ion batteries: {T}he state of the art and challenges
  of electrode materials reacting through conversion reactions. \emph{Adv.
  Mater.} \textbf{2010}, \emph{22}, E170--E192\relax
\mciteBstWouldAddEndPuncttrue
\mciteSetBstMidEndSepPunct{\mcitedefaultmidpunct}
{\mcitedefaultendpunct}{\mcitedefaultseppunct}\relax
\EndOfBibitem
\bibitem[Blanc \latin{et~al.}(2020)Blanc, Kundu, and Nazar]{Blanc_J_4_2020}
Blanc,~L.~E.; Kundu,~D.; Nazar,~L.~F. Scientific challenges for the
  implementation of {Zn}-ion batteries. \emph{Joule} \textbf{2020}, \emph{4},
  771--799\relax
\mciteBstWouldAddEndPuncttrue
\mciteSetBstMidEndSepPunct{\mcitedefaultmidpunct}
{\mcitedefaultendpunct}{\mcitedefaultseppunct}\relax
\EndOfBibitem
\bibitem[Fan \latin{et~al.}(2018)Fan, Ma, Wang, Yang, and Lu]{Fan_AM_30_2018}
Fan,~L.; Ma,~R.; Wang,~J.; Yang,~H.; Lu,~B. An ultrafast and highly stable
  potassium-organic battery. \emph{Adv. Mater.} \textbf{2018}, \emph{30},
  1805486\relax
\mciteBstWouldAddEndPuncttrue
\mciteSetBstMidEndSepPunct{\mcitedefaultmidpunct}
{\mcitedefaultendpunct}{\mcitedefaultseppunct}\relax
\EndOfBibitem
\bibitem[Pan \latin{et~al.}(2021)Pan, Liu, Yang, Li, Liu, Loh, and
  Wang]{Pan_AEM_11_2021}
Pan,~Z.; Liu,~X.; Yang,~J.; Li,~X.; Liu,~Z.; Loh,~X.~J.; Wang,~J. Aqueous
  rechargeable multivalent metal-ion batteries: advances and challenges.
  \emph{Adv. Energy Mater.} \textbf{2021}, \emph{11}, 2100608\relax
\mciteBstWouldAddEndPuncttrue
\mciteSetBstMidEndSepPunct{\mcitedefaultmidpunct}
{\mcitedefaultendpunct}{\mcitedefaultseppunct}\relax
\EndOfBibitem
\bibitem[Zhang \latin{et~al.}(2020)Zhang, Chen, Yu, Niu, Cheng, and
  Chen]{Zhang_CSR_49_2020}
Zhang,~N.; Chen,~X.; Yu,~M.; Niu,~Z.; Cheng,~F.; Chen,~J. Materials chemistry
  for rechargeable zinc-ion batteries. \emph{Chem. Soc. Rev.} \textbf{2020},
  \emph{49}, 4203--4219\relax
\mciteBstWouldAddEndPuncttrue
\mciteSetBstMidEndSepPunct{\mcitedefaultmidpunct}
{\mcitedefaultendpunct}{\mcitedefaultseppunct}\relax
\EndOfBibitem
\bibitem[Selvakumaran \latin{et~al.}(2019)Selvakumaran, Pan, Liang, and
  Cao]{Selvakumaran_JMCA_7_2019}
Selvakumaran,~D.; Pan,~A.; Liang,~S.; Cao,~G. A review on recent developments
  and challenges of cathode materials for rechargeable aqueous {Zn}-ion
  batteries. \emph{J. Mater. Chem. A} \textbf{2019}, \emph{7},
  18209--18236\relax
\mciteBstWouldAddEndPuncttrue
\mciteSetBstMidEndSepPunct{\mcitedefaultmidpunct}
{\mcitedefaultendpunct}{\mcitedefaultseppunct}\relax
\EndOfBibitem
\bibitem[Zhao \latin{et~al.}(2019)Zhao, Zhu, and Zhang]{Zhao_I_2_2019}
Zhao,~Y.; Zhu,~Y.; Zhang,~X. Challenges and perspectives for manganese-based
  oxides for advanced aqueous zinc-ion batteries. \emph{{InfoMat}}
  \textbf{2019}, \emph{2}, 237--260\relax
\mciteBstWouldAddEndPuncttrue
\mciteSetBstMidEndSepPunct{\mcitedefaultmidpunct}
{\mcitedefaultendpunct}{\mcitedefaultseppunct}\relax
\EndOfBibitem
\bibitem[Pan \latin{et~al.}(2016)Pan, Shao, Yan, Cheng, Han, Nie, Wang, Yang,
  Li, Bhattacharya, Mueller, and Liu]{Pan_NE_1_2016}
Pan,~H.; Shao,~Y.; Yan,~P.; Cheng,~Y.; Han,~K.~S.; Nie,~Z.; Wang,~C.; Yang,~J.;
  Li,~X.; Bhattacharya,~P.; Mueller,~K.~T.; Liu,~J. Reversible aqueous
  zinc/manganese oxide energy storage from conversion reactions. \emph{Nat.
  Energy} \textbf{2016}, \emph{1}, 16039\relax
\mciteBstWouldAddEndPuncttrue
\mciteSetBstMidEndSepPunct{\mcitedefaultmidpunct}
{\mcitedefaultendpunct}{\mcitedefaultseppunct}\relax
\EndOfBibitem
\bibitem[Chen \latin{et~al.}(2019)Chen, An, and Mai]{Chen_AMI_6_2019}
Chen,~L.; An,~Q.; Mai,~L. Recent advances and prospects of cathode materials
  for rechargeable aqueous zinc-ion batteries. \emph{Adv. Mater. Interfaces}
  \textbf{2019}, \emph{6}, 1900387\relax
\mciteBstWouldAddEndPuncttrue
\mciteSetBstMidEndSepPunct{\mcitedefaultmidpunct}
{\mcitedefaultendpunct}{\mcitedefaultseppunct}\relax
\EndOfBibitem
\bibitem[Jiang \latin{et~al.}(2017)Jiang, Xu, Wu, Dong, Li, and
  Kang]{Jiang_EA_229_2017}
Jiang,~B.; Xu,~C.; Wu,~C.; Dong,~L.; Li,~J.; Kang,~F. Manganese sesquioxide as
  cathode material for multivalent zinc ion battery with high capacity and long
  cycle life. \emph{Electrochim. Acta} \textbf{2017}, \emph{229},
  422--428\relax
\mciteBstWouldAddEndPuncttrue
\mciteSetBstMidEndSepPunct{\mcitedefaultmidpunct}
{\mcitedefaultendpunct}{\mcitedefaultseppunct}\relax
\EndOfBibitem
\bibitem[Wu \latin{et~al.}(2020)Wu, Housel, Kim, Sadique, Quilty, Wu, Tappero,
  Nicholas, Ehrlich, Zhu, Marschilok, Takeuchi, Bock, and
  Takeuchi]{Wu_EES_13_2020}
Wu,~D.; Housel,~L.~M.; Kim,~S.~J.; Sadique,~N.; Quilty,~C.~D.; Wu,~L.;
  Tappero,~R.; Nicholas,~S.~L.; Ehrlich,~S.; Zhu,~Y.; Marschilok,~A.~C.;
  Takeuchi,~E.~S.; Bock,~D.~C.; Takeuchi,~K.~J. Quantitative temporally and
  spatially resolved {X}-ray fluorescence microprobe characterization of the
  manganese dissolution-deposition mechanism in aqueous {Zn/$\alpha$-MnO$_2$}
  batteries. \emph{Energ. Environ. Sci.} \textbf{2020}, \emph{13},
  4322--4333\relax
\mciteBstWouldAddEndPuncttrue
\mciteSetBstMidEndSepPunct{\mcitedefaultmidpunct}
{\mcitedefaultendpunct}{\mcitedefaultseppunct}\relax
\EndOfBibitem
\bibitem[Tran \latin{et~al.}(2021)Tran, Jin, Cuisinier, Adams, and
  Ivey]{Tran_SR_11_2021}
Tran,~T. N.~T.; Jin,~S.; Cuisinier,~M.; Adams,~B.~D.; Ivey,~D.~G. Reaction
  mechanisms for electrolytic manganese dioxide in rechargeable aqueous
  zinc-ion batteries. \emph{Sci. Rep.} \textbf{2021}, \emph{11}, 20777\relax
\mciteBstWouldAddEndPuncttrue
\mciteSetBstMidEndSepPunct{\mcitedefaultmidpunct}
{\mcitedefaultendpunct}{\mcitedefaultseppunct}\relax
\EndOfBibitem
\bibitem[Rubel \latin{et~al.}(2022)Rubel, Tran, Gourley, Anand, Bommel, Adams,
  Ivey, and Higgins]{Rubel_JPCC_126_2022}
Rubel,~O.; Tran,~T. N.~T.; Gourley,~S.; Anand,~S.; Bommel,~A.~V.; Adams,~B.~D.;
  Ivey,~D.~G.; Higgins,~D. Electrochemical stability of {ZnMn$_2$O$_4$}:
  {U}nderstanding {Zn}-ion rechargeable battery capacity and degradation.
  \emph{J. Phys. Chem. C} \textbf{2022}, \emph{126}, 10957--10967\relax
\mciteBstWouldAddEndPuncttrue
\mciteSetBstMidEndSepPunct{\mcitedefaultmidpunct}
{\mcitedefaultendpunct}{\mcitedefaultseppunct}\relax
\EndOfBibitem
\bibitem[Chevrier \latin{et~al.}(2010)Chevrier, Ong, Armiento, Chan, and
  Ceder]{Chevrier_PRB_82_2010}
Chevrier,~V.~L.; Ong,~S.~P.; Armiento,~R.; Chan,~M. K.~Y.; Ceder,~G. Hybrid
  density functional calculations of redox potentials and formation energies of
  transition metal compounds. \emph{Phys. Rev. B} \textbf{2010}, \emph{82},
  075122\relax
\mciteBstWouldAddEndPuncttrue
\mciteSetBstMidEndSepPunct{\mcitedefaultmidpunct}
{\mcitedefaultendpunct}{\mcitedefaultseppunct}\relax
\EndOfBibitem
\bibitem[Aydinol \latin{et~al.}(1997)Aydinol, Kohan, Ceder, Cho, and
  Joannopoulos]{Aydinol_PRB_56_1997}
Aydinol,~M.~K.; Kohan,~A.~F.; Ceder,~G.; Cho,~K.; Joannopoulos,~J. Ab initio
  study of lithium intercalation in metal oxides and metal dichalcogenides.
  \emph{Phys. Rev. B} \textbf{1997}, \emph{56}, 1354--1365\relax
\mciteBstWouldAddEndPuncttrue
\mciteSetBstMidEndSepPunct{\mcitedefaultmidpunct}
{\mcitedefaultendpunct}{\mcitedefaultseppunct}\relax
\EndOfBibitem
\bibitem[Zhang \latin{et~al.}(2019)Zhang, Zhang, Zhao, Yao, Chen, and
  Zhou]{Zhang_AO_4_2019}
Zhang,~Z.; Zhang,~X.; Zhao,~X.; Yao,~S.; Chen,~A.; Zhou,~Z. Computational
  Screening of Layered Materials for Multivalent Ion Batteries. \emph{{ACS}
  Omega} \textbf{2019}, \emph{4}, 7822--7828\relax
\mciteBstWouldAddEndPuncttrue
\mciteSetBstMidEndSepPunct{\mcitedefaultmidpunct}
{\mcitedefaultendpunct}{\mcitedefaultseppunct}\relax
\EndOfBibitem
\bibitem[Hohenberg and Kohn(1964)Hohenberg, and Kohn]{Hohenberg_PR_136_1964}
Hohenberg,~P.; Kohn,~W. Inhomogeneous electron gas. \emph{Phys. Rev.}
  \textbf{1964}, \emph{136}, B864--B871\relax
\mciteBstWouldAddEndPuncttrue
\mciteSetBstMidEndSepPunct{\mcitedefaultmidpunct}
{\mcitedefaultendpunct}{\mcitedefaultseppunct}\relax
\EndOfBibitem
\bibitem[Kohn and Sham(1965)Kohn, and Sham]{Kohn_PR_140_1965}
Kohn,~W.; Sham,~L.~J. Self-consistent equations including exchange and
  correlation effects. \emph{Phys. Rev.} \textbf{1965}, \emph{140},
  A1133--A1138\relax
\mciteBstWouldAddEndPuncttrue
\mciteSetBstMidEndSepPunct{\mcitedefaultmidpunct}
{\mcitedefaultendpunct}{\mcitedefaultseppunct}\relax
\EndOfBibitem
\bibitem[Jain \latin{et~al.}(2013)Jain, Ong, Hautier, Chen, Richards, Dacek,
  Cholia, Gunter, Skinner, Ceder, and Persson]{Jain_AM_1_2013}
Jain,~A.; Ong,~S.~P.; Hautier,~G.; Chen,~W.; Richards,~W.~D.; Dacek,~S.;
  Cholia,~S.; Gunter,~D.; Skinner,~D.; Ceder,~G.; Persson,~K.~A. Commentary:
  {T}he {M}aterials {P}roject: {A} materials genome approach to accelerating
  materials innovation. \emph{{APL} Mater.} \textbf{2013}, \emph{1},
  011002\relax
\mciteBstWouldAddEndPuncttrue
\mciteSetBstMidEndSepPunct{\mcitedefaultmidpunct}
{\mcitedefaultendpunct}{\mcitedefaultseppunct}\relax
\EndOfBibitem
\bibitem[Ong \latin{et~al.}(2011)Ong, Chevrier, Hautier, Jain, Moore, Kim, Ma,
  and Ceder]{Ong_EES_4_2011}
Ong,~S.~P.; Chevrier,~V.~L.; Hautier,~G.; Jain,~A.; Moore,~C.; Kim,~S.; Ma,~X.;
  Ceder,~G. Voltage, stability and diffusion barrier differences between
  sodium-ion and lithium-ion intercalation materials. \emph{Energ. Environ.
  Sci.} \textbf{2011}, \emph{4}, 3680\relax
\mciteBstWouldAddEndPuncttrue
\mciteSetBstMidEndSepPunct{\mcitedefaultmidpunct}
{\mcitedefaultendpunct}{\mcitedefaultseppunct}\relax
\EndOfBibitem
\bibitem[Le \latin{et~al.}(2021)Le, Sadique, Housel, Poyraz, Takeuchi,
  Takeuchi, Marschilok, and Liu]{Le_AAMI_13_2021}
Le,~T.; Sadique,~N.; Housel,~L.~M.; Poyraz,~A.~S.; Takeuchi,~E.~S.;
  Takeuchi,~K.~J.; Marschilok,~A.~C.; Liu,~P. Discharging behavior of
  hollandite $\alpha$-{MnO}$_2$ in a hydrated zinc-ion battery. \emph{{ACS}
  Appl. Mater. Interfaces} \textbf{2021}, \emph{13}, 59937--59949\relax
\mciteBstWouldAddEndPuncttrue
\mciteSetBstMidEndSepPunct{\mcitedefaultmidpunct}
{\mcitedefaultendpunct}{\mcitedefaultseppunct}\relax
\EndOfBibitem
\bibitem[Perdew \latin{et~al.}(1996)Perdew, Burke, and
  Ernzerhof]{Perdew_PRL_77_1996}
Perdew,~J.~P.; Burke,~K.; Ernzerhof,~M. Generalized gradient approximation made
  simple. \emph{Phys. Rev. Lett.} \textbf{1996}, \emph{77}, 3865--3868\relax
\mciteBstWouldAddEndPuncttrue
\mciteSetBstMidEndSepPunct{\mcitedefaultmidpunct}
{\mcitedefaultendpunct}{\mcitedefaultseppunct}\relax
\EndOfBibitem
\bibitem[Luo \latin{et~al.}(2022)Luo, Deng, Gou, Odunmbaku, Sun, Xiao, Li, and
  Zheng]{Luo_CCL_unknown_2022}
Luo,~H.; Deng,~J.; Gou,~Q.; Odunmbaku,~O.; Sun,~K.; Xiao,~J.; Li,~M.; Zheng,~Y.
  Accelerated discovery of novel high-performance zinc-ion battery cathode
  materials by combining high-throughput screening and experiments. \emph{Chin.
  Chem. Lett.} \textbf{2022}, 107885\relax
\mciteBstWouldAddEndPuncttrue
\mciteSetBstMidEndSepPunct{\mcitedefaultmidpunct}
{\mcitedefaultendpunct}{\mcitedefaultseppunct}\relax
\EndOfBibitem
\bibitem[Kresse and Hafner(1993)Kresse, and Hafner]{Kresse_PRB_47_1993}
Kresse,~G.; Hafner,~J. Ab initio molecular dynamics for liquid metals.
  \emph{Phys. Rev. B} \textbf{1993}, \emph{47}, 558\relax
\mciteBstWouldAddEndPuncttrue
\mciteSetBstMidEndSepPunct{\mcitedefaultmidpunct}
{\mcitedefaultendpunct}{\mcitedefaultseppunct}\relax
\EndOfBibitem
\bibitem[Kresse and Furthm{\"u}ller(1996)Kresse, and
  Furthm{\"u}ller]{Kresse_CMS_6_1996}
Kresse,~G.; Furthm{\"u}ller,~J. Efficiency of ab-initio total energy
  calculations for metals and semiconductors using a plane-wave basis set.
  \emph{Comp. Mater. Sci.} \textbf{1996}, \emph{6}, 15--50\relax
\mciteBstWouldAddEndPuncttrue
\mciteSetBstMidEndSepPunct{\mcitedefaultmidpunct}
{\mcitedefaultendpunct}{\mcitedefaultseppunct}\relax
\EndOfBibitem
\bibitem[Kresse and Furthm{\"u}ller(1996)Kresse, and
  Furthm{\"u}ller]{Kresse_PRB_54_1996}
Kresse,~G.; Furthm{\"u}ller,~J. Efficient iterative schemes for ab initio
  total-energy calculations using a plane-wave basis set. \emph{Phys. Rev. B}
  \textbf{1996}, \emph{54}, 11169--11186\relax
\mciteBstWouldAddEndPuncttrue
\mciteSetBstMidEndSepPunct{\mcitedefaultmidpunct}
{\mcitedefaultendpunct}{\mcitedefaultseppunct}\relax
\EndOfBibitem
\bibitem[Grimme \latin{et~al.}(2010)Grimme, Antony, Ehrlich, and
  Krieg]{Grimme_JCP_132_2010}
Grimme,~S.; Antony,~J.; Ehrlich,~S.; Krieg,~H. A consistent and accurate ab
  initio parametrization of density functional dispersion correction ({DFT-D})
  for the 94 elements {H-Pu}. \emph{J. Chem. Phys.} \textbf{2010}, \emph{132},
  154104\relax
\mciteBstWouldAddEndPuncttrue
\mciteSetBstMidEndSepPunct{\mcitedefaultmidpunct}
{\mcitedefaultendpunct}{\mcitedefaultseppunct}\relax
\EndOfBibitem
\bibitem[Grimme \latin{et~al.}(2011)Grimme, Ehrlich, and
  Goerigk]{Grimme_JCC_32_2011}
Grimme,~S.; Ehrlich,~S.; Goerigk,~L. Effect of the damping function in
  dispersion corrected density functional theory. \emph{J. Comput. Chem.}
  \textbf{2011}, \emph{32}, 1456--1465\relax
\mciteBstWouldAddEndPuncttrue
\mciteSetBstMidEndSepPunct{\mcitedefaultmidpunct}
{\mcitedefaultendpunct}{\mcitedefaultseppunct}\relax
\EndOfBibitem
\bibitem[Kresse and Joubert(1999)Kresse, and Joubert]{Kresse_PRB_59_1999}
Kresse,~G.; Joubert,~D. From ultrasoft pseudopotentials to the projector
  augmented-wave method. \emph{Phys. Rev. B} \textbf{1999}, \emph{59},
  1758--1775\relax
\mciteBstWouldAddEndPuncttrue
\mciteSetBstMidEndSepPunct{\mcitedefaultmidpunct}
{\mcitedefaultendpunct}{\mcitedefaultseppunct}\relax
\EndOfBibitem
\bibitem[Monkhorst and Pack(1976)Monkhorst, and Pack]{Monkhorst_PRB_13_1976}
Monkhorst,~H.~J.; Pack,~J.~D. Special points for {Brillouin-zone} integrations.
  \emph{Phys. Rev. B} \textbf{1976}, \emph{13}, 5188--5192\relax
\mciteBstWouldAddEndPuncttrue
\mciteSetBstMidEndSepPunct{\mcitedefaultmidpunct}
{\mcitedefaultendpunct}{\mcitedefaultseppunct}\relax
\EndOfBibitem
\bibitem[Dudarev \latin{et~al.}(1998)Dudarev, Botton, Savrasov, Humphreys, and
  Sutton]{Dudarev_PRB_57_1998}
Dudarev,~S.~L.; Botton,~G.~A.; Savrasov,~S.~Y.; Humphreys,~C.~J.; Sutton,~A.~P.
  Electron-energy-loss spectra and the structural stability of nickel oxide: An
  {LSDA+U} study. \emph{Phys. Rev. B} \textbf{1998}, \emph{57},
  1505--1509\relax
\mciteBstWouldAddEndPuncttrue
\mciteSetBstMidEndSepPunct{\mcitedefaultmidpunct}
{\mcitedefaultendpunct}{\mcitedefaultseppunct}\relax
\EndOfBibitem
\bibitem[Capdevila-Cortada \latin{et~al.}(2016)Capdevila-Cortada, {\L}odziana,
  and L{\'{o}}pez]{CapdevilaCortada_AC_6_2016}
Capdevila-Cortada,~M.; {\L}odziana,~Z.; L{\'{o}}pez,~N. Performance of
  {DFT$+U$} Approaches in the Study of Catalytic Materials. \emph{{ACS} Catal.}
  \textbf{2016}, \emph{6}, 8370--8379\relax
\mciteBstWouldAddEndPuncttrue
\mciteSetBstMidEndSepPunct{\mcitedefaultmidpunct}
{\mcitedefaultendpunct}{\mcitedefaultseppunct}\relax
\EndOfBibitem
\bibitem[Jain \latin{et~al.}(2011)Jain, Hautier, Ong, Moore, Fischer, Persson,
  and Ceder]{Jain_PRB_84_2011}
Jain,~A.; Hautier,~G.; Ong,~S.~P.; Moore,~C.~J.; Fischer,~C.~C.;
  Persson,~K.~A.; Ceder,~G. Formation enthalpies by mixing {GGA} and
  $\text{GGA}+{U}$ calculations. \emph{Phys. Rev. B} \textbf{2011}, \emph{84},
  045115\relax
\mciteBstWouldAddEndPuncttrue
\mciteSetBstMidEndSepPunct{\mcitedefaultmidpunct}
{\mcitedefaultendpunct}{\mcitedefaultseppunct}\relax
\EndOfBibitem
\bibitem[J{\'o}nsson \latin{et~al.}(1998)J{\'o}nsson, Mills, and
  Jacobsen]{Jonsson_class-quant-dynam_1998}
J{\'o}nsson,~H.; Mills,~G.; Jacobsen,~K.~W. \emph{Classical and quantum
  dynamics in condensed phase simulations}; World Scientific, 1998; pp
  385--404\relax
\mciteBstWouldAddEndPuncttrue
\mciteSetBstMidEndSepPunct{\mcitedefaultmidpunct}
{\mcitedefaultendpunct}{\mcitedefaultseppunct}\relax
\EndOfBibitem
\bibitem[Mills \latin{et~al.}(1995)Mills, J{\'{o}}nsson, and
  Schenter]{Mills_SS_324_1995}
Mills,~G.; J{\'{o}}nsson,~H.; Schenter,~G.~K. Reversible work transition state
  theory: application to dissociative adsorption of hydrogen. \emph{Surf. Sci.}
  \textbf{1995}, \emph{324}, 305--337\relax
\mciteBstWouldAddEndPuncttrue
\mciteSetBstMidEndSepPunct{\mcitedefaultmidpunct}
{\mcitedefaultendpunct}{\mcitedefaultseppunct}\relax
\EndOfBibitem
\bibitem[Lan(2001)]{LandoltBornstein_thermo_2001}
Thermodynamic Properties of Compounds, {LiO} to {MnS$_2$}: {D}atasheet from
  {Landolt-B{\"o}rnstein} - Group {IV} Physical Chemistry - Volume {19A4}:
  ``{P}ure Substances. {P}art 4 - Compounds from {HgH{\_}g to ZnTe{\_}g}'' in
  {SpringerMaterials} (https://doi.org/10.1007/10688868{\_}10). 2001;
  \url{https://materials.springer.com/lb/docs/sm_lbs_978-3-540-45274-4_10}\relax
\mciteBstWouldAddEndPuncttrue
\mciteSetBstMidEndSepPunct{\mcitedefaultmidpunct}
{\mcitedefaultendpunct}{\mcitedefaultseppunct}\relax
\EndOfBibitem
\bibitem[Franchini \latin{et~al.}(2007)Franchini, Podloucky, Paier, Marsman,
  and Kresse]{Franchini_PRB_75_2007}
Franchini,~C.; Podloucky,~R.; Paier,~J.; Marsman,~M.; Kresse,~G. Ground-state
  properties of multivalent manganese oxides: {D}ensity functional and hybrid
  density functional calculations. \emph{Phys. Rev. B} \textbf{2007},
  \emph{75}, 195128\relax
\mciteBstWouldAddEndPuncttrue
\mciteSetBstMidEndSepPunct{\mcitedefaultmidpunct}
{\mcitedefaultendpunct}{\mcitedefaultseppunct}\relax
\EndOfBibitem
\bibitem[Eckhoff \latin{et~al.}(2020)Eckhoff, Bl{\"o}chl, and
  Behler]{Eckhoff_PRB_101_2020}
Eckhoff,~M.; Bl{\"o}chl,~P.~E.; Behler,~J. Hybrid density functional theory
  benchmark study on lithium manganese oxides. \emph{Phys. Rev. B}
  \textbf{2020}, \emph{101}, 205113\relax
\mciteBstWouldAddEndPuncttrue
\mciteSetBstMidEndSepPunct{\mcitedefaultmidpunct}
{\mcitedefaultendpunct}{\mcitedefaultseppunct}\relax
\EndOfBibitem
\bibitem[Chase(1998)]{Chase_JPCRD_Monograph9_1998}
Chase,~M.~W. {NIST-JANAF} Thermochemical Tables, 4th Edition. \emph{J. Phys.
  Chem. Ref. Data} \textbf{1998}, \emph{Monograph 9}, 1--1951\relax
\mciteBstWouldAddEndPuncttrue
\mciteSetBstMidEndSepPunct{\mcitedefaultmidpunct}
{\mcitedefaultendpunct}{\mcitedefaultseppunct}\relax
\EndOfBibitem
\bibitem[Das \latin{et~al.}(2019)Das, Tosoni, and Pacchioni]{Das_CMS_163_2019}
Das,~T.; Tosoni,~S.; Pacchioni,~G. Structural and electronic properties of bulk
  and ultrathin layers of {V$_2$O$_5$} and {MoO$_3$}. \emph{Comp. Mater. Sci.}
  \textbf{2019}, \emph{163}, 230--240\relax
\mciteBstWouldAddEndPuncttrue
\mciteSetBstMidEndSepPunct{\mcitedefaultmidpunct}
{\mcitedefaultendpunct}{\mcitedefaultseppunct}\relax
\EndOfBibitem
\bibitem[Brown \latin{et~al.}(2005)Brown, Curti, Grambow, and
  Ekberg]{brown2005chemical}
Brown,~P.~L.; Curti,~E.; Grambow,~B.; Ekberg,~C. \emph{Chemical thermodynamics
  of zirconium}; Elsevier Amsterdam, 2005; Vol.~8\relax
\mciteBstWouldAddEndPuncttrue
\mciteSetBstMidEndSepPunct{\mcitedefaultmidpunct}
{\mcitedefaultendpunct}{\mcitedefaultseppunct}\relax
\EndOfBibitem
\bibitem[Wu \latin{et~al.}(2019)Wu, Kang, Duan, and Li]{Wu_PNSMI_29_2019}
Wu,~X.; Kang,~F.; Duan,~W.; Li,~J. Density functional theory calculations: {A}
  powerful tool to simulate and design high-performance energy storage and
  conversion materials. \emph{Prog. Nat. Sci. Mater. Int.} \textbf{2019},
  \emph{29}, 247--255\relax
\mciteBstWouldAddEndPuncttrue
\mciteSetBstMidEndSepPunct{\mcitedefaultmidpunct}
{\mcitedefaultendpunct}{\mcitedefaultseppunct}\relax
\EndOfBibitem
\bibitem[Xu \latin{et~al.}(2011)Xu, Li, Du, and Kang]{Xu_ACIE_51_2011}
Xu,~C.; Li,~B.; Du,~H.; Kang,~F. Energetic zinc ion chemistry: {T}he
  rechargeable zinc ion battery. \emph{Angew. Chem. Int. Ed.} \textbf{2011},
  \emph{51}, 933--935\relax
\mciteBstWouldAddEndPuncttrue
\mciteSetBstMidEndSepPunct{\mcitedefaultmidpunct}
{\mcitedefaultendpunct}{\mcitedefaultseppunct}\relax
\EndOfBibitem
\bibitem[Alfaruqi \latin{et~al.}(2015)Alfaruqi, Mathew, Gim, Kim, Song, Baboo,
  Choi, and Kim]{Alfaruqi_CM_27_2015}
Alfaruqi,~M.~H.; Mathew,~V.; Gim,~J.; Kim,~S.; Song,~J.; Baboo,~J.~P.;
  Choi,~S.~H.; Kim,~J. Electrochemically Induced Structural Transformation in a
  $\gamma$-{MnO}$_2$ Cathode of a High Capacity Zinc-Ion Battery System.
  \emph{Chem. Mater.} \textbf{2015}, \emph{27}, 3609--3620\relax
\mciteBstWouldAddEndPuncttrue
\mciteSetBstMidEndSepPunct{\mcitedefaultmidpunct}
{\mcitedefaultendpunct}{\mcitedefaultseppunct}\relax
\EndOfBibitem
\bibitem[Kundu \latin{et~al.}(2016)Kundu, Adams, Duffort, Vajargah, and
  Nazar]{Kundu_NE_1_2016}
Kundu,~D.; Adams,~B.~D.; Duffort,~V.; Vajargah,~S.~H.; Nazar,~L.~F. A
  high-capacity and long-life aqueous rechargeable zinc battery using a metal
  oxide intercalation cathode. \emph{Nat. Energy} \textbf{2016}, \emph{1},
  4599\relax
\mciteBstWouldAddEndPuncttrue
\mciteSetBstMidEndSepPunct{\mcitedefaultmidpunct}
{\mcitedefaultendpunct}{\mcitedefaultseppunct}\relax
\EndOfBibitem
\bibitem[Song \latin{et~al.}(2018)Song, Tan, Chao, and Fan]{Song_AFM_28_2018}
Song,~M.; Tan,~H.; Chao,~D.; Fan,~H.~J. Recent Advances in Zn-Ion Batteries.
  \emph{Adv. Funct. Mater.} \textbf{2018}, \emph{28}, 1802564\relax
\mciteBstWouldAddEndPuncttrue
\mciteSetBstMidEndSepPunct{\mcitedefaultmidpunct}
{\mcitedefaultendpunct}{\mcitedefaultseppunct}\relax
\EndOfBibitem
\bibitem[Yang \latin{et~al.}(2020)Yang, Li, Ma, Hong, and
  Wang]{Yang_JMCA_8_2020}
Yang,~G.; Li,~Q.; Ma,~K.; Hong,~C.; Wang,~C. The degradation mechanism of
  vanadium oxide-based aqueous zinc-ion batteries. \emph{J. Mater. Chem. A}
  \textbf{2020}, \emph{8}, 8084--8095\relax
\mciteBstWouldAddEndPuncttrue
\mciteSetBstMidEndSepPunct{\mcitedefaultmidpunct}
{\mcitedefaultendpunct}{\mcitedefaultseppunct}\relax
\EndOfBibitem
\bibitem[Byeon \latin{et~al.}(2021)Byeon, Hong, Bae, Shin, Choi, and
  Chung]{Byeon_NC_12_2021}
Byeon,~P.; Hong,~Y.; Bae,~H.~B.; Shin,~J.; Choi,~J.~W.; Chung,~S.-Y.
  Atomic-scale unveiling of multiphase evolution during hydrated {Zn}-ion
  insertion in vanadium oxide. \emph{Nat. Commun.} \textbf{2021}, \emph{12},
  4599\relax
\mciteBstWouldAddEndPuncttrue
\mciteSetBstMidEndSepPunct{\mcitedefaultmidpunct}
{\mcitedefaultendpunct}{\mcitedefaultseppunct}\relax
\EndOfBibitem
\bibitem[Haas \latin{et~al.}(2009)Haas, Tran, and Blaha]{Haas_PRB_79_2009}
Haas,~P.; Tran,~F.; Blaha,~P. Calculation of the lattice constant of solids
  with semilocal functionals. \emph{Phys. Rev. B} \textbf{2009}, \emph{79},
  085104\relax
\mciteBstWouldAddEndPuncttrue
\mciteSetBstMidEndSepPunct{\mcitedefaultmidpunct}
{\mcitedefaultendpunct}{\mcitedefaultseppunct}\relax
\EndOfBibitem
\bibitem[Vil(2016)]{Villars_SpringerMaterials_2016}
{PAULING FILE} Multinaries Edition in {SpringerMaterials}. 2016;
  \url{https://materials.springer.com}, Copyright 2016 Springer-Verlag Berlin
  Heidelberg {\&} Material Phases Data System (MPDS), Switzerland {\&} National
  Institute for Materials Science (NIMS), Japan\relax
\mciteBstWouldAddEndPuncttrue
\mciteSetBstMidEndSepPunct{\mcitedefaultmidpunct}
{\mcitedefaultendpunct}{\mcitedefaultseppunct}\relax
\EndOfBibitem
\bibitem[Wang \latin{et~al.}(2006)Wang, Maxisch, and Ceder]{Wang_PRB_73_2006}
Wang,~L.; Maxisch,~T.; Ceder,~G. Oxidation energies of transition metal oxides
  within the $\mathrm{GGA}+\mathrm{U}$ framework. \emph{Phys. Rev. B}
  \textbf{2006}, \emph{73}, 195107\relax
\mciteBstWouldAddEndPuncttrue
\mciteSetBstMidEndSepPunct{\mcitedefaultmidpunct}
{\mcitedefaultendpunct}{\mcitedefaultseppunct}\relax
\EndOfBibitem
\bibitem[Urban \latin{et~al.}(2016)Urban, Seo, and Ceder]{Urban_nCM_2_2016}
Urban,~A.; Seo,~D.-H.; Ceder,~G. Computational understanding of {Li}-ion
  batteries. \emph{{npj} Comput. Mater.} \textbf{2016}, \emph{2}, 16002\relax
\mciteBstWouldAddEndPuncttrue
\mciteSetBstMidEndSepPunct{\mcitedefaultmidpunct}
{\mcitedefaultendpunct}{\mcitedefaultseppunct}\relax
\EndOfBibitem
\bibitem[Wu \latin{et~al.}(2017)Wu, Xiang, Peng, Wu, Li, Tang, Song, Liu, He,
  and Wu]{Wu_JMCA_5_2017}
Wu,~X.; Xiang,~Y.; Peng,~Q.; Wu,~X.; Li,~Y.; Tang,~F.; Song,~R.; Liu,~Z.;
  He,~Z.; Wu,~X. Green-low-cost rechargeable aqueous zinc-ion batteries using
  hollow porous spinel {ZnMn$_2$O$_4$} as the cathode material. \emph{J. Mater.
  Chem. A} \textbf{2017}, \emph{5}, 17990--17997\relax
\mciteBstWouldAddEndPuncttrue
\mciteSetBstMidEndSepPunct{\mcitedefaultmidpunct}
{\mcitedefaultendpunct}{\mcitedefaultseppunct}\relax
\EndOfBibitem
\bibitem[Tang \latin{et~al.}(2022)Tang, Chen, Lyu, and
  Chen]{Tang_EMA_2022_2022}
Tang,~Z.; Chen,~W.; Lyu,~Z.; Chen,~Q. Size-Dependent Reaction Mechanism of
  {$\lambda$-MnO$_2$} Particles as Cathodes in Aqueous Zinc-Ion Batteries.
  \emph{Energy Mater. Adv.} \textbf{2022}, \emph{2022}, 9765710\relax
\mciteBstWouldAddEndPuncttrue
\mciteSetBstMidEndSepPunct{\mcitedefaultmidpunct}
{\mcitedefaultendpunct}{\mcitedefaultseppunct}\relax
\EndOfBibitem
\bibitem[Alfaruqi \latin{et~al.}(2015)Alfaruqi, Gim, Kim, Song, Jo, Kim,
  Mathew, and Kim]{Alfaruqi_JPS_288_2015}
Alfaruqi,~M.~H.; Gim,~J.; Kim,~S.; Song,~J.; Jo,~J.; Kim,~S.; Mathew,~V.;
  Kim,~J. Enhanced reversible divalent zinc storage in a structurally stable
  $\alpha$-{MnO}$_2$ nanorod electrode. \emph{J. Power Sources} \textbf{2015},
  \emph{288}, 320--327\relax
\mciteBstWouldAddEndPuncttrue
\mciteSetBstMidEndSepPunct{\mcitedefaultmidpunct}
{\mcitedefaultendpunct}{\mcitedefaultseppunct}\relax
\EndOfBibitem
\bibitem[Lee \latin{et~al.}(2015)Lee, Lee, Kim, Chung, Cho, and
  Oh]{Lee_CC_51_2015}
Lee,~B.; Lee,~H.~R.; Kim,~H.; Chung,~K.~Y.; Cho,~B.~W.; Oh,~S.~H. Elucidating
  the intercalation mechanism of zinc ions into {$\alpha$-MnO$_2$} for
  rechargeable zinc batteries. \emph{Chem. Commun.} \textbf{2015}, \emph{51},
  9265--9268\relax
\mciteBstWouldAddEndPuncttrue
\mciteSetBstMidEndSepPunct{\mcitedefaultmidpunct}
{\mcitedefaultendpunct}{\mcitedefaultseppunct}\relax
\EndOfBibitem
\bibitem[Pang \latin{et~al.}(2021)Pang, He, Yu, Yang, Zhao, Fu, Xing, Tian,
  Luo, and Wei]{Pang_ASS_538_2021}
Pang,~Q.; He,~W.; Yu,~X.; Yang,~S.; Zhao,~H.; Fu,~Y.; Xing,~M.; Tian,~Y.;
  Luo,~X.; Wei,~Y. Aluminium pre-intercalated orthorhombic {V$_2$O$_5$} as
  high-performance cathode material for aqueous zinc-ion batteries. \emph{Appl.
  Surf. Sci.} \textbf{2021}, \emph{538}, 148043\relax
\mciteBstWouldAddEndPuncttrue
\mciteSetBstMidEndSepPunct{\mcitedefaultmidpunct}
{\mcitedefaultendpunct}{\mcitedefaultseppunct}\relax
\EndOfBibitem
\bibitem[Hu \latin{et~al.}(2017)Hu, Yan, Zhu, Wang, Wei, Li, Zhou, Li, Chen,
  and Mai]{Hu_AAMI_9_2017}
Hu,~P.; Yan,~M.; Zhu,~T.; Wang,~X.; Wei,~X.; Li,~J.; Zhou,~L.; Li,~Z.;
  Chen,~L.; Mai,~L. {Zn/V$_2$O$_5$} Aqueous Hybrid-Ion Battery with High
  Voltage Platform and Long Cycle Life. \emph{{ACS} Appl. Mater. Interfaces}
  \textbf{2017}, \emph{9}, 42717--42722\relax
\mciteBstWouldAddEndPuncttrue
\mciteSetBstMidEndSepPunct{\mcitedefaultmidpunct}
{\mcitedefaultendpunct}{\mcitedefaultseppunct}\relax
\EndOfBibitem
\bibitem[Lee \latin{et~al.}(2014)Lee, Yoon, Lee, Chung, Cho, and
  Oh]{Lee_SR_4_2014}
Lee,~B.; Yoon,~C.~S.; Lee,~H.~R.; Chung,~K.~Y.; Cho,~B.~W.; Oh,~S.~H.
  Electrochemically-induced reversible transition from the tunneled to layered
  polymorphs of manganese dioxide. \emph{Sci. Rep.} \textbf{2014}, \emph{4},
  6066\relax
\mciteBstWouldAddEndPuncttrue
\mciteSetBstMidEndSepPunct{\mcitedefaultmidpunct}
{\mcitedefaultendpunct}{\mcitedefaultseppunct}\relax
\EndOfBibitem
\bibitem[Moon \latin{et~al.}(2021)Moon, Ha, Park, Lee, Kwon, Lim, Lee, Kim,
  Choi, Choi, and Lee]{Moon_AS_8_2021}
Moon,~H.; Ha,~K.-H.; Park,~Y.; Lee,~J.; Kwon,~M.-S.; Lim,~J.; Lee,~M.-H.;
  Kim,~D.-H.; Choi,~J.~H.; Choi,~J.-H.; Lee,~K.~T. Direct proof of the
  reversible dissolution/deposition of {Mn$^{2+}$/Mn$^{4+}$} for mild-acid
  {Zn-MnO$_2$} batteries with porous carbon interlayers. \emph{Adv. Sci.}
  \textbf{2021}, \emph{8}, 2003714\relax
\mciteBstWouldAddEndPuncttrue
\mciteSetBstMidEndSepPunct{\mcitedefaultmidpunct}
{\mcitedefaultendpunct}{\mcitedefaultseppunct}\relax
\EndOfBibitem
\bibitem[Chen \latin{et~al.}(2022)Chen, Dai, Xiao, Yang, Cai, Xu, Fan, and
  Bao]{Chen_AM_34_2022}
Chen,~H.; Dai,~C.; Xiao,~F.; Yang,~Q.; Cai,~S.; Xu,~M.; Fan,~H.~J.; Bao,~S.-J.
  Reunderstanding the reaction mechanism of aqueous {Zn-Mn} batteries with
  sulfate electrolytes: {R}ole of the zinc sulfate hydroxide. \emph{Adv.
  Mater.} \textbf{2022}, \emph{34}, 2109092\relax
\mciteBstWouldAddEndPuncttrue
\mciteSetBstMidEndSepPunct{\mcitedefaultmidpunct}
{\mcitedefaultendpunct}{\mcitedefaultseppunct}\relax
\EndOfBibitem
\bibitem[Jiao \latin{et~al.}(2020)Jiao, Kang, Berry-Gair, McColl, Li, Dong,
  Jiang, Wang, Cor{\`{a}}, Brett, He, and Parkin]{Jiao_JMCA_8_2020}
Jiao,~Y.; Kang,~L.; Berry-Gair,~J.; McColl,~K.; Li,~J.; Dong,~H.; Jiang,~H.;
  Wang,~R.; Cor{\`{a}},~F.; Brett,~D. J.~L.; He,~G.; Parkin,~I.~P. Enabling
  stable {MnO$_2$} matrix for aqueous zinc-ion battery cathodes. \emph{J.
  Mater. Chem. A} \textbf{2020}, \emph{8}, 22075--22082\relax
\mciteBstWouldAddEndPuncttrue
\mciteSetBstMidEndSepPunct{\mcitedefaultmidpunct}
{\mcitedefaultendpunct}{\mcitedefaultseppunct}\relax
\EndOfBibitem
\bibitem[Gautam \latin{et~al.}(2015)Gautam, Canepa, Malik, Liu, Persson, and
  Ceder]{Gautam_CC_51_2015}
Gautam,~G.~S.; Canepa,~P.; Malik,~R.; Liu,~M.; Persson,~K.; Ceder,~G.
  First-principles evaluation of multi-valent cation insertion into
  orthorhombic {V$_2$O$_5$}. \emph{Chem. Commun.} \textbf{2015}, \emph{51},
  13619--13622\relax
\mciteBstWouldAddEndPuncttrue
\mciteSetBstMidEndSepPunct{\mcitedefaultmidpunct}
{\mcitedefaultendpunct}{\mcitedefaultseppunct}\relax
\EndOfBibitem
\bibitem[Sandagiripathira \latin{et~al.}(2022)Sandagiripathira, Moghaddasi,
  Shepard, and Smeu]{Sandagiripathira_PCCP_24_2022}
Sandagiripathira,~K.; Moghaddasi,~M.~A.; Shepard,~R.; Smeu,~M. Investigating
  the role of structural water on the electrochemical properties of
  {$\alpha$-V$_2$O$_5$} through density functional theory. \emph{Phys. Chem.
  Chem. Phys.} \textbf{2022}, \emph{24}, 24271--24280\relax
\mciteBstWouldAddEndPuncttrue
\mciteSetBstMidEndSepPunct{\mcitedefaultmidpunct}
{\mcitedefaultendpunct}{\mcitedefaultseppunct}\relax
\EndOfBibitem
\bibitem[Horrocks \latin{et~al.}(2013)Horrocks, Likely, Velazquez, and
  Banerjee]{Horrocks_JMCA_1_2013}
Horrocks,~G.~A.; Likely,~M.~F.; Velazquez,~J.~M.; Banerjee,~S. Finite size
  effects on the structural progression induced by lithiation of {V$_2$O$_5$}:
  a combined diffraction and {R}aman spectroscopy study. \emph{J. Mater. Chem.
  A} \textbf{2013}, \emph{1}, 15265\relax
\mciteBstWouldAddEndPuncttrue
\mciteSetBstMidEndSepPunct{\mcitedefaultmidpunct}
{\mcitedefaultendpunct}{\mcitedefaultseppunct}\relax
\EndOfBibitem
\bibitem[Gautam \latin{et~al.}(2015)Gautam, Canepa, Abdellahi, Urban, Malik,
  and Ceder]{Gautam_CM_27_2015}
Gautam,~G.~S.; Canepa,~P.; Abdellahi,~A.; Urban,~A.; Malik,~R.; Ceder,~G. The
  intercalation phase diagram of {Mg} in {V$_2$O$_5$} from first-principles.
  \emph{Chem. Mater.} \textbf{2015}, \emph{27}, 3733--3742\relax
\mciteBstWouldAddEndPuncttrue
\mciteSetBstMidEndSepPunct{\mcitedefaultmidpunct}
{\mcitedefaultendpunct}{\mcitedefaultseppunct}\relax
\EndOfBibitem
\bibitem[Putro \latin{et~al.}(2020)Putro, Alfaruqi, Islam, Kim, Park, Lee,
  Hwang, Sun, and Kim]{Putro_EA_345_2020}
Putro,~D.~Y.; Alfaruqi,~M.~H.; Islam,~S.; Kim,~S.; Park,~S.; Lee,~S.;
  Hwang,~J.-Y.; Sun,~Y.-K.; Kim,~J. Quasi-solid-state zinc-ion battery based on
  {$\alpha$-MnO$_2$} cathode with husk-like morphology. \emph{Electrochim.
  Acta} \textbf{2020}, \emph{345}, 136189\relax
\mciteBstWouldAddEndPuncttrue
\mciteSetBstMidEndSepPunct{\mcitedefaultmidpunct}
{\mcitedefaultendpunct}{\mcitedefaultseppunct}\relax
\EndOfBibitem
\bibitem[Rong \latin{et~al.}(2015)Rong, Malik, Canepa, Sai~Gautam, Liu, Jain,
  Persson, and Ceder]{Rong_CM_27_2015}
Rong,~Z.; Malik,~R.; Canepa,~P.; Sai~Gautam,~G.; Liu,~M.; Jain,~A.;
  Persson,~K.; Ceder,~G. Materials design rules for multivalent ion mobility in
  intercalation structures. \emph{Chem. Mater.} \textbf{2015}, \emph{27},
  6016--6021\relax
\mciteBstWouldAddEndPuncttrue
\mciteSetBstMidEndSepPunct{\mcitedefaultmidpunct}
{\mcitedefaultendpunct}{\mcitedefaultseppunct}\relax
\EndOfBibitem
\bibitem[Yi \latin{et~al.}(2021)Yi, Qiu, Qu, Liu, Zhang, and
  Zhu]{Yi_CCR_446_2021}
Yi,~T.-F.; Qiu,~L.; Qu,~J.-P.; Liu,~H.; Zhang,~J.-H.; Zhu,~Y.-R. Towards
  high-performance cathodes: {D}esign and energy storage mechanism of vanadium
  oxides-based materials for aqueous {Zn}-ion batteries. \emph{Coord. Chem.
  Rev.} \textbf{2021}, \emph{446}, 214124\relax
\mciteBstWouldAddEndPuncttrue
\mciteSetBstMidEndSepPunct{\mcitedefaultmidpunct}
{\mcitedefaultendpunct}{\mcitedefaultseppunct}\relax
\EndOfBibitem
\bibitem[Slater(1964)]{Slater_JCP_41_1964}
Slater,~J.~C. Atomic Radii in Crystals. \emph{J. Chem. Phys.} \textbf{1964},
  \emph{41}, 3199--3204\relax
\mciteBstWouldAddEndPuncttrue
\mciteSetBstMidEndSepPunct{\mcitedefaultmidpunct}
{\mcitedefaultendpunct}{\mcitedefaultseppunct}\relax
\EndOfBibitem
\bibitem[Downs and Hall-Wallace(2003)Downs, and Hall-Wallace]{Downs_AM_88_2003}
Downs,~R.~T.; Hall-Wallace,~M. The american mineralogist crystal structure
  database. \emph{Am. Mineral.} \textbf{2003}, \emph{88}, 247--250\relax
\mciteBstWouldAddEndPuncttrue
\mciteSetBstMidEndSepPunct{\mcitedefaultmidpunct}
{\mcitedefaultendpunct}{\mcitedefaultseppunct}\relax
\EndOfBibitem
\bibitem[Gra{\v{z}}ulis \latin{et~al.}(2009)Gra{\v{z}}ulis, Chateigner, Downs,
  Yokochi, Quir{\'{o}}s, Lutterotti, Manakova, Butkus, Moeck, and
  Bail]{Grazulis_JAC_42_2009}
Gra{\v{z}}ulis,~S.; Chateigner,~D.; Downs,~R.~T.; Yokochi,~A. F.~T.;
  Quir{\'{o}}s,~M.; Lutterotti,~L.; Manakova,~E.; Butkus,~J.; Moeck,~P.;
  Bail,~A.~L. Crystallography Open Database---an open-access collection of
  crystal structures. \emph{J. Appl. Crystallogr.} \textbf{2009}, \emph{42},
  726--729\relax
\mciteBstWouldAddEndPuncttrue
\mciteSetBstMidEndSepPunct{\mcitedefaultmidpunct}
{\mcitedefaultendpunct}{\mcitedefaultseppunct}\relax
\EndOfBibitem
\bibitem[Gra{\v{z}}ulis \latin{et~al.}(2011)Gra{\v{z}}ulis,
  Da{\v{s}}kevi{\v{c}}, Merkys, Chateigner, Lutterotti, Quir{\'{o}}s,
  Serebryanaya, Moeck, Downs, and Bail]{Grazulis_NAR_40_2011}
Gra{\v{z}}ulis,~S.; Da{\v{s}}kevi{\v{c}},~A.; Merkys,~A.; Chateigner,~D.;
  Lutterotti,~L.; Quir{\'{o}}s,~M.; Serebryanaya,~N.~R.; Moeck,~P.;
  Downs,~R.~T.; Bail,~A.~L. Crystallography Open Database ({COD}): an
  open-access collection of crystal structures and platform for world-wide
  collaboration. \emph{Nucleic Acids Res.} \textbf{2011}, \emph{40},
  D420--D427\relax
\mciteBstWouldAddEndPuncttrue
\mciteSetBstMidEndSepPunct{\mcitedefaultmidpunct}
{\mcitedefaultendpunct}{\mcitedefaultseppunct}\relax
\EndOfBibitem
\bibitem[Shin \latin{et~al.}(2019)Shin, Choi, Lee, Jung, and
  Choi]{Shin_AEM_9_2019}
Shin,~J.; Choi,~D.~S.; Lee,~H.~J.; Jung,~Y.; Choi,~J.~W. Hydrated Intercalation
  for High-Performance Aqueous Zinc Ion Batteries. \emph{Adv. Energy Mater.}
  \textbf{2019}, \emph{9}, 1900083\relax
\mciteBstWouldAddEndPuncttrue
\mciteSetBstMidEndSepPunct{\mcitedefaultmidpunct}
{\mcitedefaultendpunct}{\mcitedefaultseppunct}\relax
\EndOfBibitem
\bibitem[Adams \latin{et~al.}(2020)Adams, Brown, Cuisinier, and
  Jin]{Adams_patent_layered_2020}
Adams,~B.~D.; Brown,~R.~D.; Cuisinier,~M.; Jin,~S. Layered electrode materials
  and methods for rechargeable zinc batteries. 2020; patent US 2020/0395606
  A1\relax
\mciteBstWouldAddEndPuncttrue
\mciteSetBstMidEndSepPunct{\mcitedefaultmidpunct}
{\mcitedefaultendpunct}{\mcitedefaultseppunct}\relax
\EndOfBibitem
\bibitem[Hou \latin{et~al.}(2020)Hou, Tan, Gao, Li, Lu, and
  Zhang]{Hou_JMC_8_2020}
Hou,~Z.; Tan,~H.; Gao,~Y.; Li,~M.; Lu,~Z.; Zhang,~B. Tailoring desolvation
  kinetics enables stable zinc metal anodes. \emph{J. Mater. Chem.A}
  \textbf{2020}, \emph{8}, 19367--19374\relax
\mciteBstWouldAddEndPuncttrue
\mciteSetBstMidEndSepPunct{\mcitedefaultmidpunct}
{\mcitedefaultendpunct}{\mcitedefaultseppunct}\relax
\EndOfBibitem
\bibitem[Cau{\"{e}}t \latin{et~al.}(2010)Cau{\"{e}}t, Bogatko, Weare, Fulton,
  Schenter, and Bylaska]{Cauet_JCP_132_2010}
Cau{\"{e}}t,~E.; Bogatko,~S.; Weare,~J.~H.; Fulton,~J.~L.; Schenter,~G.~K.;
  Bylaska,~E.~J. Structure and dynamics of the hydration shells of the
  {Zn$^{2+}$} ion from \textit{ab initio} molecular dynamics and combined
  \textit{ab initio} and classical molecular dynamics simulations. \emph{J.
  Chem. Phys.} \textbf{2010}, \emph{132}, 194502\relax
\mciteBstWouldAddEndPuncttrue
\mciteSetBstMidEndSepPunct{\mcitedefaultmidpunct}
{\mcitedefaultendpunct}{\mcitedefaultseppunct}\relax
\EndOfBibitem
\bibitem[Brady \latin{et~al.}(2019)Brady, Tallman, Takeuchi, Marschilok,
  Takeuchi, and Liu]{Brady_JPCC_123_2019}
Brady,~A.~B.; Tallman,~K.~R.; Takeuchi,~E.~S.; Marschilok,~A.~C.;
  Takeuchi,~K.~J.; Liu,~P. Transition metal substitution of hollandite
  $\alpha$-{MnO}$_2$: enhanced potential and structural stability on lithiation
  from first-principles calculation. \emph{J. Phys. Chem. C} \textbf{2019},
  \emph{123}, 25042--25051\relax
\mciteBstWouldAddEndPuncttrue
\mciteSetBstMidEndSepPunct{\mcitedefaultmidpunct}
{\mcitedefaultendpunct}{\mcitedefaultseppunct}\relax
\EndOfBibitem
\bibitem[Post \latin{et~al.}(2003)Post, Heaney, Dreele, and
  Hanson]{Post_AM_88_2003}
Post,~J.~E.; Heaney,~P.~J.; Dreele,~R. B.~V.; Hanson,~J.~C. Neutron and
  temperature-resolved synchrotron {X}-ray powder diffraction study of
  akagan{\'{e}}ite. \emph{Am. Mineral.} \textbf{2003}, \emph{88},
  782--788\relax
\mciteBstWouldAddEndPuncttrue
\mciteSetBstMidEndSepPunct{\mcitedefaultmidpunct}
{\mcitedefaultendpunct}{\mcitedefaultseppunct}\relax
\EndOfBibitem
\bibitem[Fernandez-Martinez \latin{et~al.}(2010)Fernandez-Martinez, Timon,
  Roman-Ross, Cuello, Daniels, and Ayora]{FernandezMartinez_AM_95_2010}
Fernandez-Martinez,~A.; Timon,~V.; Roman-Ross,~G.; Cuello,~G.~J.;
  Daniels,~J.~E.; Ayora,~C. The structure of schwertmannite, a nanocrystalline
  iron oxyhydroxysulfate. \emph{Am. Mineral.} \textbf{2010}, \emph{95},
  1312--1322\relax
\mciteBstWouldAddEndPuncttrue
\mciteSetBstMidEndSepPunct{\mcitedefaultmidpunct}
{\mcitedefaultendpunct}{\mcitedefaultseppunct}\relax
\EndOfBibitem
\bibitem[Zhang \latin{et~al.}(2017)Zhang, Pelliccione, Brady, Guo, Smith, Liu,
  Marschilok, Takeuchi, and Takeuchi]{Zhang_CM_29_2017}
Zhang,~Y.; Pelliccione,~C.~J.; Brady,~A.~B.; Guo,~H.; Smith,~P.~F.; Liu,~P.;
  Marschilok,~A.~C.; Takeuchi,~K.~J.; Takeuchi,~E.~S. Probing the {Li}
  insertion mechanism of {ZnFe$_2$O$_4$} in {Li}-ion batteries: a combined
  x-ray diffraction, extended x-ray absorption fine structure, and density
  functional theory study. \emph{Chem. Mater.} \textbf{2017}, \emph{29},
  4282--4292\relax
\mciteBstWouldAddEndPuncttrue
\mciteSetBstMidEndSepPunct{\mcitedefaultmidpunct}
{\mcitedefaultendpunct}{\mcitedefaultseppunct}\relax
\EndOfBibitem
\bibitem[Liu \latin{et~al.}(2017)Liu, Hao, Xu, Mou, Dong, Jiang, Kang, Wu,
  Jiang, and Kang]{Liu_CC_53_2017}
Liu,~W.; Hao,~J.; Xu,~C.; Mou,~J.; Dong,~L.; Jiang,~F.; Kang,~Z.; Wu,~J.;
  Jiang,~B.; Kang,~F. Investigation of zinc ion storage of transition metal
  oxides, sulfides, and borides in zinc ion battery systems. \emph{Chem.
  Commun.} \textbf{2017}, \emph{53}, 6872--6874\relax
\mciteBstWouldAddEndPuncttrue
\mciteSetBstMidEndSepPunct{\mcitedefaultmidpunct}
{\mcitedefaultendpunct}{\mcitedefaultseppunct}\relax
\EndOfBibitem
\bibitem[Lee \latin{et~al.}(2020)Lee, Xiong, Wang, and Xue]{Lee_SM_5_2020}
Lee,~W. S.~V.; Xiong,~T.; Wang,~X.; Xue,~J. Unraveling {MoS$_2$} and transition
  metal dichalcogenides as functional zinc-ion battery cathode: a perspective.
  \emph{Small Methods} \textbf{2020}, \emph{5}, 2000815\relax
\mciteBstWouldAddEndPuncttrue
\mciteSetBstMidEndSepPunct{\mcitedefaultmidpunct}
{\mcitedefaultendpunct}{\mcitedefaultseppunct}\relax
\EndOfBibitem
\bibitem[Li \latin{et~al.}(2019)Li, Zhang, Zhou, Wu, Zhang, and
  Zhang]{Li_AN_13_2019}
Li,~Y.; Zhang,~R.; Zhou,~W.; Wu,~X.; Zhang,~H.; Zhang,~J. Hierarchical
  {MoS$_2$} hollow architectures with abundant {Mo} vacancies for efficient
  sodium storage. \emph{{ACS} Nano} \textbf{2019}, \emph{13}, 5533--5540\relax
\mciteBstWouldAddEndPuncttrue
\mciteSetBstMidEndSepPunct{\mcitedefaultmidpunct}
{\mcitedefaultendpunct}{\mcitedefaultseppunct}\relax
\EndOfBibitem
\bibitem[Liu \latin{et~al.}(2022)Liu, Yang, Chen, Chen, Zhang, Zeng, Lei,
  Huang, Li, and Peng]{Liu_EA_410_2022}
Liu,~L.; Yang,~W.; Chen,~H.; Chen,~X.; Zhang,~K.; Zeng,~Q.; Lei,~S.; Huang,~J.;
  Li,~S.; Peng,~S. High zinc-ion intercalation reaction activity of {MoS}$_2$
  cathode based on regulation of thermodynamic metastability and interlayer
  water. \emph{Electrochim. Acta} \textbf{2022}, \emph{410}, 140016\relax
\mciteBstWouldAddEndPuncttrue
\mciteSetBstMidEndSepPunct{\mcitedefaultmidpunct}
{\mcitedefaultendpunct}{\mcitedefaultseppunct}\relax
\EndOfBibitem
\bibitem[Liu \latin{et~al.}(2012)Liu, Zhong, Huang, Mao, Zhu, and
  Huang]{Liu_AN_6_2012}
Liu,~X.~H.; Zhong,~L.; Huang,~S.; Mao,~S.~X.; Zhu,~T.; Huang,~J.~Y.
  Size-dependent fracture of silicon nanoparticles during lithiation.
  \emph{{ACS} Nano} \textbf{2012}, \emph{6}, 1522--1531\relax
\mciteBstWouldAddEndPuncttrue
\mciteSetBstMidEndSepPunct{\mcitedefaultmidpunct}
{\mcitedefaultendpunct}{\mcitedefaultseppunct}\relax
\EndOfBibitem
\bibitem[{de Biasi} \latin{et~al.}(2017){de Biasi}, Lieser, Dr{\"a}ger, Indris,
  Rana, Schumacher, M{\"o}nig, Ehrenberg, Binder, and
  Ge{\ss}wein]{Biasi_JPS_362_2017}
{de Biasi},~L.; Lieser,~G.; Dr{\"a}ger,~C.; Indris,~S.; Rana,~J.;
  Schumacher,~G.; M{\"o}nig,~R.; Ehrenberg,~H.; Binder,~J.~R.; Ge{\ss}wein,~H.
  {LiCaFeF$_6$}: {A} zero-strain cathode material for use in {Li}-ion
  batteries. \emph{J. Power Sources} \textbf{2017}, \emph{362}, 192--201\relax
\mciteBstWouldAddEndPuncttrue
\mciteSetBstMidEndSepPunct{\mcitedefaultmidpunct}
{\mcitedefaultendpunct}{\mcitedefaultseppunct}\relax
\EndOfBibitem
\bibitem[Moorhead-Rosenberg \latin{et~al.}(2013)Moorhead-Rosenberg, Harrison,
  Turner, and Manthiram]{MoorheadRosenberg_IC_52_2013}
Moorhead-Rosenberg,~Z.; Harrison,~K.~L.; Turner,~T.; Manthiram,~A. A rapid
  microwave-assisted solvothermal approach to lower-valent transition metal
  oxides. \emph{Inorg. Chem.} \textbf{2013}, \emph{52}, 13087--13093\relax
\mciteBstWouldAddEndPuncttrue
\mciteSetBstMidEndSepPunct{\mcitedefaultmidpunct}
{\mcitedefaultendpunct}{\mcitedefaultseppunct}\relax
\EndOfBibitem
\bibitem[Wu \latin{et~al.}(2019)Wu, Zhu, Qin, and Huang]{Wu_JMCA_7_2019}
Wu,~T.; Zhu,~K.; Qin,~C.; Huang,~K. Unraveling the role of structural water in
  bilayer {V$_2$O$_5$} during {Zn$^{2+}$}-intercalation: insights from {DFT}
  calculations. \emph{J. Mater. Chem. A} \textbf{2019}, \emph{7},
  5612--5620\relax
\mciteBstWouldAddEndPuncttrue
\mciteSetBstMidEndSepPunct{\mcitedefaultmidpunct}
{\mcitedefaultendpunct}{\mcitedefaultseppunct}\relax
\EndOfBibitem
\bibitem[Ong \latin{et~al.}(2008)Ong, Wang, Kang, and Ceder]{Ong_CM_20_2008}
Ong,~S.~P.; Wang,~L.; Kang,~B.; Ceder,~G. {Li-Fe-P-O$_2$} phase diagram from
  first principles calculations. \emph{Chem. Mater.} \textbf{2008}, \emph{20},
  1798--1807\relax
\mciteBstWouldAddEndPuncttrue
\mciteSetBstMidEndSepPunct{\mcitedefaultmidpunct}
{\mcitedefaultendpunct}{\mcitedefaultseppunct}\relax
\EndOfBibitem
\bibitem[Predel(1997)]{Predel_Mo-O_phase-diagram_1997}
Predel,~B. {Mo-O} (Molybdenum-Oxygen): Datasheet from {L}andolt-{B}{\"o}rnstein
  - Group {IV} Physical Chemistry - Volume {5H}: ``{Li-Mg -- Nd-Zr}'' in
  {SpringerMaterials} (https://doi.org/10.1007/10522884{\_}2076). 1997;
  \url{https://materials.springer.com/lb/docs/sm_lbs_978-3-540-68538-8_2076},
  Copyright 1997 Springer-Verlag Berlin Heidelberg\relax
\mciteBstWouldAddEndPuncttrue
\mciteSetBstMidEndSepPunct{\mcitedefaultmidpunct}
{\mcitedefaultendpunct}{\mcitedefaultseppunct}\relax
\EndOfBibitem
\bibitem[Reichelt \latin{et~al.}(2000)Reichelt, Weber, S{\"o}hnel, and
  D{\"a}britz]{Reichelt_ZAAC_626_2000}
Reichelt,~W.; Weber,~T.; S{\"o}hnel,~T.; D{\"a}britz,~S. {M}ischkristallbildung
  im {S}ystem {CuMoO}$_4$/{ZnMoO}$_4$. \emph{Z. Anorg. Allg. Chem.}
  \textbf{2000}, \emph{626}, 2020--2027\relax
\mciteBstWouldAddEndPuncttrue
\mciteSetBstMidEndSepPunct{\mcitedefaultmidpunct}
{\mcitedefaultendpunct}{\mcitedefaultseppunct}\relax
\EndOfBibitem
\bibitem[S{\"o}hnel \latin{et~al.}(1997)S{\"o}hnel, Reichelt, and
  Oppermann]{Soehnel_ZAAC_623_1997}
S{\"o}hnel,~T.; Reichelt,~W.; Oppermann,~H. {Z}um {S}ystem {Zn/Mo/O}. {II}.
  {C}hemischer {T}ransport tern{\"a}rer {Z}inkmolybdate. \emph{Z. Anorg. Allg.
  Chem.} \textbf{1997}, \emph{623}, 1190--1200\relax
\mciteBstWouldAddEndPuncttrue
\mciteSetBstMidEndSepPunct{\mcitedefaultmidpunct}
{\mcitedefaultendpunct}{\mcitedefaultseppunct}\relax
\EndOfBibitem
\bibitem[Kubaschewski and Schmid-Fetzer(2007)Kubaschewski, and
  Schmid-Fetzer]{Kubaschewski_Fe-O_phase-diagram_2007}
Kubaschewski,~O.; Schmid-Fetzer,~R. Temperature - composition phase diagram of
  the {Fe-O} system: Datasheet from {MSI Eureka in SpringerMaterials}
  (https://materials.springer.com/msi/phase-diagram/docs/sm{\_}msi{\_}r{\_}10{\_}011481{\_}02{\_}full{\_}LnkDia2).
  2007;
  \url{https://materials.springer.com/msi/phase-diagram/docs/sm_msi_r_10_011481_02_full_LnkDia2},
  Copyright 2007 MSI, Materials Science International Services GmbH,
  Stuttgart\relax
\mciteBstWouldAddEndPuncttrue
\mciteSetBstMidEndSepPunct{\mcitedefaultmidpunct}
{\mcitedefaultendpunct}{\mcitedefaultseppunct}\relax
\EndOfBibitem
\bibitem[Hu \latin{et~al.}(2016)Hu, Kim, Yang, Yang, Meng, Zhang, and
  Mao]{Hu_N_534_2016}
Hu,~Q.; Kim,~D.~Y.; Yang,~W.; Yang,~L.; Meng,~Y.; Zhang,~L.; Mao,~H.-K.
  {FeO}$_2$ and {FeOOH} under deep lower-mantle conditions and {E}arth's
  oxygen-hydrogen cycles. \emph{Nature} \textbf{2016}, \emph{534},
  241--244\relax
\mciteBstWouldAddEndPuncttrue
\mciteSetBstMidEndSepPunct{\mcitedefaultmidpunct}
{\mcitedefaultendpunct}{\mcitedefaultseppunct}\relax
\EndOfBibitem
\bibitem[Watson \latin{et~al.}(2016)Watson, Chang, Dang, Gotcu-Freis, Khvan,
  Markus, Schuster, and Strafela]{MSIEureka2016:sm_msi_r_10_029498_01}
Watson,~A.; Chang,~K.; Dang,~S.; Gotcu-Freis,~P.; Khvan,~A.; Markus,~T.;
  Schuster,~E.; Strafela,~M. {Co-Li-O} ternary phase diagram evaluation:
  {D}atasheet from {MSI} {E}ureka in {S}pringer{M}aterials (dataset {ID}:
  sm{\_}msi{\_}r{\_}10{\_}029498{\_}01). 2016;
  \url{https://materials.springer.com/msi/docs/sm_msi_r_10_029498_01}\relax
\mciteBstWouldAddEndPuncttrue
\mciteSetBstMidEndSepPunct{\mcitedefaultmidpunct}
{\mcitedefaultendpunct}{\mcitedefaultseppunct}\relax
\EndOfBibitem
\bibitem[Chang(2013)]{Chang_PhD_thesis_2013}
Chang,~K. Phase equilibria, thermodynamic and electrochemical properties
  ofcathodes in lithium ion batteries based on the {Li-(Co, Ni)-O} system.
  Ph.D.\ thesis, Rheinisch-{W}estf{\"a}lischen {T}echnischen {H}ochschule
  {A}achen, 2013\relax
\mciteBstWouldAddEndPuncttrue
\mciteSetBstMidEndSepPunct{\mcitedefaultmidpunct}
{\mcitedefaultendpunct}{\mcitedefaultseppunct}\relax
\EndOfBibitem
\bibitem[Bischoff \latin{et~al.}(2020)Bischoff, Fitz, Burns, Bauer, Gentischer,
  Birke, Henning, and Biro]{Bischoff_JES_167_2020}
Bischoff,~C.~F.; Fitz,~O.~S.; Burns,~J.; Bauer,~M.; Gentischer,~H.;
  Birke,~K.~P.; Henning,~H.-M.; Biro,~D. Revealing the local {pH} value changes
  of acidic aqueous zinc ion batteries with a manganese dioxide electrode
  during cycling. \emph{J. Electrochem. Soc.} \textbf{2020}, \emph{167},
  020545\relax
\mciteBstWouldAddEndPuncttrue
\mciteSetBstMidEndSepPunct{\mcitedefaultmidpunct}
{\mcitedefaultendpunct}{\mcitedefaultseppunct}\relax
\EndOfBibitem
\bibitem[Chamoun \latin{et~al.}(2018)Chamoun, Brant, Tai, Karlsson, and
  Nor{\'{e}}us]{Chamoun_ESM_15_2018}
Chamoun,~M.; Brant,~W.~R.; Tai,~C.-W.; Karlsson,~G.; Nor{\'{e}}us,~D.
  Rechargeability of aqueous sulfate {Zn/MnO}$_2$ batteries enhanced by
  accessible {Mn$^{2+}$} ions. \emph{Energy Stor. Mater.} \textbf{2018},
  \emph{15}, 351--360\relax
\mciteBstWouldAddEndPuncttrue
\mciteSetBstMidEndSepPunct{\mcitedefaultmidpunct}
{\mcitedefaultendpunct}{\mcitedefaultseppunct}\relax
\EndOfBibitem
\bibitem[Persson \latin{et~al.}(2012)Persson, Waldwick, Lazic, and
  Ceder]{Persson_PRB_85_2012}
Persson,~K.~A.; Waldwick,~B.; Lazic,~P.; Ceder,~G. Prediction of solid-aqueous
  equilibria: Scheme to combine first-principles calculations of solids with
  experimental aqueous states. \emph{Phys. Rev. B} \textbf{2012}, \emph{85},
  235438\relax
\mciteBstWouldAddEndPuncttrue
\mciteSetBstMidEndSepPunct{\mcitedefaultmidpunct}
{\mcitedefaultendpunct}{\mcitedefaultseppunct}\relax
\EndOfBibitem
\bibitem[Zeng \latin{et~al.}(2015)Zeng, Chan, Zhao, Kubal, Fan, and
  Greeley]{Zeng_JPCC_119_2015}
Zeng,~Z.; Chan,~M. K.~Y.; Zhao,~Z.-J.; Kubal,~J.; Fan,~D.; Greeley,~J. Towards
  first principles-based prediction of highly accurate electrochemical
  {P}ourbaix diagrams. \emph{J. Phys. Chem. C} \textbf{2015}, \emph{119},
  18177--18187\relax
\mciteBstWouldAddEndPuncttrue
\mciteSetBstMidEndSepPunct{\mcitedefaultmidpunct}
{\mcitedefaultendpunct}{\mcitedefaultseppunct}\relax
\EndOfBibitem
\bibitem[Wang \latin{et~al.}(2020)Wang, Guo, Montoya, and
  N{\o}rskov]{Wang_nCM_6_2020}
Wang,~Z.; Guo,~X.; Montoya,~J.; N{\o}rskov,~J.~K. Predicting aqueous stability
  of solid with computed {P}ourbaix diagram using {SCAN} functional. \emph{npj
  Comput. Mater.} \textbf{2020}, \emph{6}, 160\relax
\mciteBstWouldAddEndPuncttrue
\mciteSetBstMidEndSepPunct{\mcitedefaultmidpunct}
{\mcitedefaultendpunct}{\mcitedefaultseppunct}\relax
\EndOfBibitem
\bibitem[Bale \latin{et~al.}(2016)Bale, B{\'{e}}lisle, Chartrand, Decterov,
  Eriksson, Gheribi, Hack, Jung, Kang, Melan{\c{c}}on, Pelton, Petersen,
  Robelin, Sangster, Spencer, and Ende]{Bale_C_54_2016}
Bale,~C.~W.; B{\'{e}}lisle,~E.; Chartrand,~P.; Decterov,~S.~A.; Eriksson,~G.;
  Gheribi,~A.~E.; Hack,~K.; Jung,~I.-H.; Kang,~Y.-B.; Melan{\c{c}}on,~J.;
  Pelton,~A.~D.; Petersen,~S.; Robelin,~C.; Sangster,~J.; Spencer,~P.;
  Ende,~M.-A.~V. {FactSage} thermochemical software and databases, 2010--2016.
  \emph{Calphad} \textbf{2016}, \emph{54}, 35--53\relax
\mciteBstWouldAddEndPuncttrue
\mciteSetBstMidEndSepPunct{\mcitedefaultmidpunct}
{\mcitedefaultendpunct}{\mcitedefaultseppunct}\relax
\EndOfBibitem
\bibitem[Zhang \latin{et~al.}(2018)Zhang, Dong, Jia, Bian, Wang, Qiu, Xu, Liu,
  Jiao, and Cheng]{Zhang_AEL_3_2018}
Zhang,~N.; Dong,~Y.; Jia,~M.; Bian,~X.; Wang,~Y.; Qiu,~M.; Xu,~J.; Liu,~Y.;
  Jiao,~L.; Cheng,~F. Rechargeable aqueous {Zn-V$_2$O$_5$} battery with high
  energy density and long cycle life. \emph{ACS Energy Lett.} \textbf{2018},
  \emph{3}, 1366--1372\relax
\mciteBstWouldAddEndPuncttrue
\mciteSetBstMidEndSepPunct{\mcitedefaultmidpunct}
{\mcitedefaultendpunct}{\mcitedefaultseppunct}\relax
\EndOfBibitem
\bibitem[Anand \latin{et~al.}(2023)Anand, Miliante, Storm, Adams, Higgins, and
  Rubel]{Zenodo_10.5281/zenodo.7796317}
Anand,~S.; Miliante,~C.~M.; Storm,~G.; Adams,~B.~D.; Higgins,~D.; Rubel,~O.
  Computational screening of cathode materials for {Zn}-ion rechargeable
  batteries (supporting files). \url{https://zenodo.org/record/7796317}, 2023;
  \url{https://zenodo.org/record/7796317}, {Z}enado file repository, doi:
  10.5281/zenodo.7796317\relax
\mciteBstWouldAddEndPuncttrue
\mciteSetBstMidEndSepPunct{\mcitedefaultmidpunct}
{\mcitedefaultendpunct}{\mcitedefaultseppunct}\relax
\EndOfBibitem
\end{mcitethebibliography}

\end{document}